\documentclass[hyper]{prop2015}
\usepackage[english]{babel}

\usepackage{amsmath,amssymb,amsthm}
\usepackage{amsfonts}
\usepackage{mathrsfs}
\usepackage{bbm}
\usepackage[all,knot,cmtip]{xy}

\usepackage{tikz}
\usetikzlibrary{arrows}
\usetikzlibrary{patterns}
\usetikzlibrary{decorations.markings}

\newcommand{\der}[1]{\frac{\partial}{\partial #1}} 
\newcommand{\ewith}{{~~~\mbox{with}~~~}}
\newcommand{\efor}{{~~~\mbox{for}~~~}}
\newcommand{\eand}{{~~~\mbox{and}~~~}}
\newcommand{\myxymatrix}[1]{\vcenter{\vbox{\xymatrix{#1}}}}

\newcommand{\doublearrow}{\mathrel{\substack{\longrightarrow\\[-0.6ex]
                      \longrightarrow}}}
\newcommand{\triplearrow}{\mathrel{\substack{\longrightarrow\\[-0.6ex]
                      \longrightarrow \\[-0.6ex]
                      \longrightarrow}}}
\newcommand{\quadarrow}{\mathrel{\substack{\longrightarrow\\[-0.6ex]
                      \longrightarrow \\[-0.6ex]
                      \longrightarrow\\[-0.6ex]
                      \longrightarrow}}}

\newcommand{\de}{\mathrm{e}}
\newcommand{\di}{\mathrm{i}}
\newcommand{\dd}{\mathrm{d}}
\newcommand{\id}{\mathrm{id}}
\newcommand{\im}{\mathrm{im}}
\newcommand{\inthom}{\underline{\mathsf{hom}}}
\newcommand{\unit}{\mathbbm{1}}
\newcommand{\acton}{\vartriangleright}

\newcommand{\frF}{\mathfrak{F}}
\newcommand{\frg}{\mathfrak{g}}
\newcommand{\frh}{\mathfrak{h}}

\newcommand{\IC}{\mathbbm{C}}
\newcommand{\IN}{\mathbbm{N}}
\newcommand{\IP}{\mathbbm{P}}
\newcommand{\IR}{\mathbbm{R}}
\newcommand{\IZ}{\mathbbm{Z}}

\newcommand{\sfa}{\mathsf{a}}
\newcommand{\sfA}{\mathsf{A}}
\newcommand{\sfB}{\mathsf{B}}
\newcommand{\sfCE}{\mathsf{CE}}
\newcommand{\sfG}{\mathsf{G}}
\newcommand{\sfd}{\mathsf{d}}
\newcommand{\sff}{\mathsf{f}}
\newcommand{\sfs}{\mathsf{s}}
\newcommand{\sft}{\mathsf{t}}
\newcommand{\sfFun}{\mathsf{Fun}}
\newcommand{\sfGrp}{\mathsf{Grp}}
\newcommand{\sfH}{\mathsf{H}}
\newcommand{\sfhom}{\mathsf{hom}}
\newcommand{\sfL}{\mathsf{L}}
\newcommand{\sfMfd}{\mathsf{Mfd}}
\newcommand{\sfSMfd}{\mathsf{SMfd}}
\newcommand{\sfSurSub}{\mathsf{SurSub}}
\newcommand{\sfsSMfd}{\mathsf{sSMfd}}
\newcommand{\sfSet}{\mathsf{Set}}
\newcommand{\sfSL}{\mathsf{SL}}
\newcommand{\sfSp}{\mathsf{Sp}}
\newcommand{\sfsSet}{\mathsf{sSet}}
\newcommand{\sfTop}{\mathsf{Top}}
\newcommand{\sfV}{\mathsf{V}}
\newcommand{\sfLie}{\mathsf{Lie}}

\newcommand{\calC}{\mathcal{C}}
\newcommand{\calN}{\mathcal{N}}

\newcommand{\scrC}{\mathscr{C}}
\newcommand{\scrE}{\mathscr{E}}
\newcommand{\scrG}{\mathscr{G}}
\newcommand{\scrS}{\mathscr{S}}

\newcommand{\scrX}{\mathscr{X}}
\newcommand{\scrO}{\mathscr{O}}
\newcommand{\scrZ}{\mathscr{Z}}

\category{Proceedings}
\title{$L_\infty$-Algebras, the BV Formalism, and Classical Fields}
\subtitle{\href{http://www.maths.dur.ac.uk/lms/109/index.html}{LMS/EPSRC Durham Symposium on Higher Structures in M-Theory}}
\author[Branislav Jur\v co]{Branislav Jur\v co\inst{a}}
\author[Tommaso Macrelli]{Tommaso Macrelli\inst{b}}
\author[Lorenzo Raspollini]{Lorenzo Raspollini\inst{b}}
\author[Christian S\"amann]{Christian S\"amann\inst{c}}
\author[Martin Wolf]{Martin Wolf\inst{b,}\footnote{Corresponding author e-mail:~\href{mailto:m.wolf@surrey.ac.uk}{\textsf{m.wolf@surrey.ac.uk}}}}
\address[1]{Charles University, Faculty of Mathematics and Physics, Mathematical Institute, Prague 186 75, Czech Republic}
\address[2]{Department of Mathematics, University of Surrey, Guildford GU2 7XH, United Kingdom; DMUS--MP--19--03}
\address[3]{Maxwell Institute for Mathematical Sciences and Department of Mathematics, Heriot--Watt University, Edinburgh EH14 4AS, United Kingdom; EMPG--19--08}
\shortauthors{B.~Jur\v co, T.~Macrelli, L.~Raspollini, C.~S{\"a}mann, M.~Wolf}
\begin{acknowledgements}
We would like to thank the participants of the \href{http://www.maths.dur.ac.uk/lms/109/index.html}{LMS/EPSRC Durham Symposium on Higher Structures in M-Theory} and of the workshop \href{https://ims.nus.edu.sg/events/2018/wstring/index.php}{String and M-Theory: the New Geometry of the 21st Century} for fruitful conversations. B.J.~was supported by the GA{\v C}R Grant 18-07776S. T.M.~and L.R.~are both partially supported by the EPSRC grant EP/N509772.
\end{acknowledgements}
\begin{abstract}
We summarise some of our recent works on $L_\infty$-algebras and quasi-groups with regard to higher principal bundles and their applications in twistor theory and gauge theory. In particular, after a lightning review of $L_\infty$-algebras, we discuss their Maurer--Cartan theory and explain that any classical field theory admitting an action can be reformulated in this context with the help of the Batalin--Vilkovisky formalism. As examples, we explore higher Chern--Simons theory and Yang--Mills theory. We also explain how these ideas can be combined with those of twistor theory to formulate maximally superconformal gauge theories in four and six dimensions by means of $L_\infty$-quasi-isomorphisms, and we propose a twistor space action.
\end{abstract}
\shortabstract

\begin{document}

\maketitle

\section{$L_\infty$-algebras}

$L_\infty$-algebras \cite{Zwiebach:1992ie,Stasheff:1992bb,Lada:1992wc,Lada:1994mn} are most straightforwardly introduced by means of $Q$-manifolds \cite{Alexandrov:1995kv,Kontsevich:1997vb,Severa:2001aa} and we shall follow this approach in this article. To set up the stage, we shall provide a few mathematical tools first. See e.g.~\cite{Cattaneo:2010re,Fairon:1512.02810} for details.

\subsection{$Q$-manifolds}\label{sec:Qmfd}

A {\it commutative differential graded algebra} is an associative unital commutative algebra $\sfA$ which is both a $\IZ$-graded algebra and a differential algebra so that all structures are compatible. 

In particular, the $\IZ$-grading implies that there is a decomposition $\sfA=\bigoplus_{k\in\IZ} \sfA_k$ and non-zero elements of $\sfA_k$ are called homogeneous and of degree~$k\in\IZ$. Furthermore, the product $\sfA\times\sfA\to\sfA$ is graded commutative,
\begin{equation}
 a_1a_2=(-1)^{|a_1||a_2|} a_2a_1
\end{equation}
for $a_{1,2}\in\sfA$ of homogeneous degrees $|a_{1,2}|\in\IZ$. Being differential means that $\sfA$ is equipped with differential derivations $\dd_k:\sfA_k\to\sfA_{k+1}$ of homogeneous degree~$1$. Concretely, the $\dd_k$ obey $\dd_{k+1}\circ \dd_k=0$ and
\begin{equation}
 \dd_k (a_1a_2)=(\dd_k a_1)a_2+(-1)^{|a_1|} a_1 (\dd_k a_2)
\end{equation}
for $a_{1,2}\in\sfA$ and $a_1$ of homogeneous degree~$|a_1|\in\IZ$. For the sake of brevity, we denote the $\dd_k$ collectively by $\dd$ and write $(\sfA,\dd)$ for a differential graded algebra.

A {\it morphism} $f\,:\,(\sfA,\dd)\to (\sfA',\dd')$ between two differential graded algebras $(\sfA,\dd)$ and $(\sfA',\dd')$ is a collection $f$ of degree $0$ maps $f_k:\sfA_k\to\sfA'_k$ which respect the differentials $f_{k+1}\circ\dd_k=\dd'_k\circ f_k$ for all $k\in\IZ$.

The prime example of a differential graded algebra is the de Rham complex $(\Omega^\bullet(X),\dd)$ on a smooth manifold $X$.

In the following, we shall need the degree-shift operation and dualisation which are defined as follows. For any $\IZ$-graded vector space $\sfV$ we define the degree shift by $l\in\IZ$ according to $\sfV[l]=\bigoplus_{k\in\IZ} (\sfV[l])_k$ with $(\sfV[l])_k:=\sfV_{k+l}$. Moreover, for the (vector space) dual $\sfV^*$ of $\sfV$, we have $(\sfV^*)_k:=(\sfV_{-k})^*$.

To motivate the notion of a $Q$-manifold, let us recall the following fact: differential forms $\Omega^\bullet(X)$ on a $d$-dimensional smooth manifold $X$ can be understood as the smooth functions $\scrC^\infty(T[1]X)$ on the degree-shifted tangent bundle $T[1]X$ of $X$. Indeed, working locally with coordinates $x^i$, $i=1,\ldots,d$, on $X$ and coordinates $\xi^i$ up the fibres of $T[1]X$, functions on $T[1]X$ are polynomials in $\xi^i$, that is, $f(x,\xi)=f_0(x)+\xi^i f_i(x)+\frac12 \xi^i\xi^j f_{ij}(x)+\cdots\in \scrC^\infty(T[1]X)$. The identification of $\xi^i$ with $\dd x^i$ amounts to $\scrC^\infty(T[1]X)\cong\Omega^\bullet(X)$. In addition, the de Rham differential $\dd$ corresponds to the vector field $Q=\xi^i\der{x^i}$ under this identification. The manifold $T[1]X$ together with the degree~1 vector field $Q$ form what is known as a $Q$-manifold. 

The proper definition of a $Q$-manifold requires the somewhat heavier machinery of locally ringed spaces which we recall here for the reader's convenience. A {\it ringed space} $X$ is a pair $(|X|,\scrS_X)$ where $|X|$ is a topological space and $\scrS_X$ a sheaf of rings on $|X|$ called the {\it structure sheaf} of $X$. A {\it locally ringed space} is then a ringed space $(|X|,\scrS_X)$ such that all stalks of $\scrS_X$ are local rings, that is, they have unique maximal ideals. 

A {\it morphism} $(|X|,\scrS_X)\to (|X'|,\scrS_{X'})$ of locally ringed spaces is a pair $(\phi,\phi^\sharp)$ where $\phi:|X|\to |X'|$ is a morphism of topological spaces and $\phi^\sharp\,:\,\scrS_{X'}\to \phi_*\scrS_X$ a comorphism of local rings i.e.~a map that respects the maximal ideals. Here, $\phi_*\scrS_X$ is the zeroth direct image of $\scrS_X$ under $\phi$ i.e.~for any open subset $U'$ of $|X'|$ there is a comorphism $\phi^\sharp_{U'}:\scrS_{X'}|_{U'}\to\scrS_X|_{\phi^{-1}({U'})}$. If the structure sheaves carry extra structure such as a $\IZ$-grading, then the morphism is assumed to respect this structure. 

For instance, an ordinary smooth manifold can be defined as a locally ringed space $(|X|,\scrS_X)$ for $|X|$ a topological manifold such that for each $x\in |X|$ there is an open neighbourhood $U\ni x$ and an isomorphism of locally ringed spaces $(U,\scrS_X|_U)\cong (U',\scrC^\infty_{U'})$ where $\scrC^\infty_{U'}$ is the sheaf of smooth functions on an open set $U'\subseteq\IR^d$. The stalk of $\scrS_{X}$ at a point $x\in|X|$ is the set of all germs of smooth functions at $x\in|X|$, and the maximal ideal of the stalk are the functions that vanish at $x\in|X|$. Furthermore, if $f:|X|\to|X'|$ is a continuous function between two topological manifolds $|X|$ and $|X'|$ for two manifolds $(|X|,\scrS_X)$ and $(|X'|,\scrS_{X'})$ and if there is a comorphism $\Phi\,:\,\scrS_{X'}\to \phi_*\scrS_X$ of local rings, then $\phi$ must also be smooth and $\Phi=\phi^\sharp$.

With this in mind, a {\it smooth $\IZ$-graded manifold} is a locally ringed space $X=(|X|,\scrS_X)$ for $|X|$ a topological manifold such that for each $x\in |X|$ there is an open neighbourhood $U\ni x$ and an isomorphism of locally ringed spaces $(U,\scrS_X|_U)\cong(U',\mbox{$\bigodot$}^\bullet\scrE^*_{U'}\otimes \scrC^\infty_{U'})$ where $U'\subseteq E$ is open for $E$ a Frech\'et space, $\scrC^\infty_{U'}$ is the sheaf of smooth functions on $U'$, and $\scrE_{U'}$ is a locally free $\IZ$-graded sheaf of $\scrC^\infty_{U'}$-modules on $U'$. We shall write $\scrC^\infty(X):=\Gamma(|X|,\scrS_X)$ to denote the global functions on $X$.

It can be shown \cite{JSTOR:1998201,Bonavolonta:2012fh} that any smooth $\IZ$-graded manifold must take the form of a vector bundle over an ordinary smooth manifold with the typical fibre being a $\IZ$-graded vector space. This is called {\it globally split}\footnote{Note that by definition, $Q$-manifolds are locally split.} and essentially due to the existence of a partition of unity and the fact that any smooth $\IZ$-manifold can be smoothly deformed into said vector bundle form. Note, however, that complex $\IZ$-graded manifolds are not necessarily globally split. We shall mostly be working in the real setting and hence often drop the prefix `smooth' in the following.

A {\it vector field} $V$ on a $\IZ$-graded manifold $X$ is simply a graded derivation $V:\scrC^\infty(X)\rightarrow \scrC^\infty(X)$. Specifically, for homogeneous $V$ of degree~$|V|\in\IZ$ and homogeneous $f,g\in\scrC^\infty(X)$, we have the graded Leibniz rule
\begin{equation}
V(fg)=V(f)g+(-1)^{|V|\,|f|}f\,V(g)~.
\end{equation}
The tangent bundle $TX$ of a $\IZ$-graded manifold $X$ is then simply defined to be the disjoint union of the tangent spaces which in turn are the vector spaces of derivations as in the ordinary case. Furthermore, {\it differential forms} can be defined by setting $\Omega^\bullet(X):=\scrC^\infty(T[1]X)$ upon recalling our above discussion.

We now have introduced all the necessary mathematical background to give the definition of a $Q$-manifold. A {\it $Q$-manifold} \cite{Alexandrov:1995kv,Kontsevich:1997vb,Severa:2001aa} is a $\IZ$-graded manifold $X$ equipped with a homogeneous degree 1~vector field such that $[Q,Q]=2Q^2=0$ where $[-,-]$ is the graded Lie bracket on the sheaf of vector fields on $X$. In addition, the pair $(\scrC^\infty(X),Q)$ forms a differential graded algebra.

\subsection{$L_\infty$-algebras}\label{sec:Linfty}

To begin with, consider a $\IZ$-graded manifold concentrated (i.e.~non-trivial) only in degree 1.\footnote{Here we mean that the coordinate ring is generated by degree $1$ coordinates.} Such a manifold is necessarily of the from $\frg[1]$ for $\frg$ an ordinary (real) vector space. Now, let $\xi^\alpha$ be local coordinates. The most general degree 1~vector field $Q$ is of the form 
\begin{equation}\label{eq:QLieAlg}
 Q:=-\tfrac12\xi^\alpha\xi^\beta {f_{\alpha\beta}}^\gamma\der{\xi^\gamma}~,
\end{equation}
where the ${f_{\alpha\beta}}^\gamma$ are constants. It is straightforward to check that $Q^2=0$ is equivalent to requiring the constants ${f_{\alpha\beta}}^\gamma$ to satisfy the Jacobi identity. Thus, $(\scrC^\infty(\frg[1]),Q)$ can be identified with the Chevalley--Eilenberg algebra $\sfCE(\frg):=(\bigwedge^\bullet\frg^*,\dd_{\sfCE})$ of a Lie algebra $(\frg,[-,-])$ with $[-,-]$ the Lie bracket.

Generalising the above, for a $Q$-manifold $X$ concentrated in degrees $1,\ldots,n$ we declare the pair $(\scrC^\infty(X),Q)$ to be the Chevalley--Eilenberg algebra $\sfCE(\sfL)$ of an {\it $n$-term $L_\infty$-algebra} $(\sfL,\mu_i)$ over $\IR$ with $\mu_i$, $i=1,\ldots,n$, being the {\it higher} brackets generalising the Lie bracket. Indeed, such a $Q$-manifold is necessarily of the form $\sfL[1]$ for a $\IZ$-graded vector space $\sfL=\bigoplus_{k=-n}^0 \sfL_k$ and, letting $\xi^\alpha$ be local coordinates of degree $|\xi^\alpha|\in\{1,\ldots,n\}$ on $\sfL[1]$, the vector field $Q$ given in~\eqref{eq:QLieAlg} generalises to 
\begin{equation}\label{eq:QVectorGen}
 Q:=\sum_{i=1}^n\frac{(-1)^{\frac12i(i+1)}}{i!}\xi^{\alpha_1}\cdots\xi^{\alpha_i}{f_{\alpha_1\cdots\alpha_i}}^\beta\der{\xi^\beta}~.
\end{equation}
The ${f_{\alpha_1\cdots\alpha_i}}^\beta$ are constants again but not all of them are non-zero due to the requirement of $Q$ being of degree 1. The constants ${f_{\alpha_1\cdots\alpha_i}}^\beta$ encode multilinear totally graded antisymmetric maps $\mu_i:\sfL\times\cdots\times\sfL\to\sfL$ of degree $2-i$. Indeed, letting $\tau_\alpha$ be a basis of $\sfL$ with $|\tau_\alpha|=-|\xi_\alpha|+1\in\{-n,\ldots,0\}$, we may write
\begin{equation}
 \mu_i(\tau_{\alpha_1},\ldots,\tau_{\alpha_i}):={f_{\alpha_1\cdots\alpha_i}}^\beta\tau_\beta~.
\end{equation}
The condition $Q^2=0$ amounts to the {\it higher} or {\it homotopy Jacobi identities} 
\begin{subequations}\label{eq:HJI}
\begin{equation}
\begin{aligned}
 &\sum_{j+k=i}\sum_{\sigma\in{\rm Sh}(j;i)}\chi(\sigma;\ell_1,\ldots,\ell_{i})(-1)^{k}\,\times\\
 &\kern.5cm\times\mu_{k+1}(\mu_j(\ell_{\sigma(1)},\ldots,\ell_{\sigma(j)}),\ell_{\sigma(j+1)},\ldots,\ell_{\sigma(i)})\ =\ 0
 \end{aligned}
\end{equation}
for $\ell_1,\ldots,\ell_{i}\in \sfL$ as a straightforward but lengthy calculation shows. Here, the sum over $\sigma$ is taken over all $(j;i)$ {\it shuffles} which consist of permutations $\sigma$ of $\{1,\ldots,i\}$ such that the first $j$ and the last $i-j$ images of $\sigma$ are ordered: $\sigma(1)<\cdots<\sigma(j)$ and $\sigma(j+1)<\cdots<\sigma(i)$. In addition, $\chi(\sigma;\ell_1,\ldots,\ell_i)$ is the {\it graded Koszul sign} defined implicitly by
\begin{equation}
 \ell_1\wedge \ldots \wedge \ell_i=\chi(\sigma;\ell_1,\ldots,\ell_i)\,\ell_{\sigma(1)}\wedge \ldots \wedge \ell_{\sigma(i)}~.
\end{equation}
\end{subequations}
In particular, for $i=1$ we find that $\mu_1$ is a differential, for $i=2$ we find that $\mu_1$ is a derivation with respect to $\mu_2$, for $i=3$ we find a generalisation of the Jacobi identity for the 2-bracket $\mu_2$, and so on.

Upon recalling the fact that any $\IZ$-graded manifold is a fibration over an ordinary manifold with the typical fibre being a $\IZ$-graded vector space, we call $(\scrC^\infty(X),Q)$ for a $Q$-manifold $X$ fibred over a point the Chevalley--Eilenberg algebra $\sfCE(\sfL)$ of an $L_\infty$-algebra $(\sfL,\mu_i)$ over $\IR$ for $i\in\IN$ and $\sfL=\bigoplus_{k\in\IZ}\sfL_k$. This extension to a $\IZ$-grading is needed when talking about the  Batalin--Vilkovisky formalism later on. 

To complete our brief exposition on $L_\infty$-algebras, we wish to introduce two more ingredients: inner products on $L_\infty$-algebras and morphisms between $L_\infty$-algebras. We shall start with the former.

Inner product $L_\infty$-algebras, also known as {\it cyclic} $L_\infty$-algebras are $L_\infty$-algebras  that come equipped with a bilinear non-degenerate graded symmetric pairing $\langle -,-\rangle\,:\,\sfL\times\sfL\to\IR$ which is cyclic in the sense of
\begin{equation}
\begin{aligned}
 &\langle\ell_1,\mu_i(\ell_2,\ldots,\ell_{i+1})\rangle=\\
 &\kern.5cm=(-1)^{i+i(|\ell_1|+|\ell_{i+1}|)+|\ell_{i+1}|\sum_{j=1}^{i}|\ell_j|}\times\\
 &\kern1.5cm\times\langle\ell_{i+1},\mu_i(\ell_1,\ldots,\ell_{i})\rangle
 \end{aligned}
\end{equation}
for all $i\in\IN$ for homogeneous $\ell_1,\ldots,\ell_{i+1}\in\sfL$ with $|\ell_i|_\sfL$ the $L_\infty$-degree of $\ell_i\in\sfL$. The inner product may carry a degree itself. In the $Q$-manifold picture, the inner product corresponds to a symplectic form, and the cyclicity is encoded in the requirement of the vector field $Q$ to be symplectic with respect to the symplectic form.

Before moving on to morphisms, let us point out that given a commutative differential graded algebra $(\sfA,\dd)$ and an $L_\infty$-algebra $(\sfL,\mu_i)$ we can always form their tensor product which again comes with an $L_\infty$-structure. Explicitly, we have
\begin{subequations}\label{eq:LinftyExtension}
\begin{equation}
 \!\!\hat{~\sfL}:=\bigoplus_{k\in\IZ} (\sfA\otimes \sfL)_k\ewith (\sfA\otimes \sfL)_k:=\bigoplus_{i+j=k}~\sfA_i\otimes \sfL_j
\end{equation}
so that the homogeneous degree in $\!\!\hat{~\sfL}$ is given by $|a\otimes \ell|:=|a|+|\ell|$ for homogeneous $a\in \sfA$ and $\ell\in \sfL$. The higher products $\hat\mu_i$ on $\!\!\hat{~\sfL}$ read as 
\begin{equation}
\begin{aligned}
 \hat\mu_1(a_1 \otimes \ell_1)&:=\dd a_1 \otimes \ell_1+(-1)^{|a_1|}a_1\otimes \mu_1(\ell_1)~,\\[5pt]
 \hat\mu_i(a_1\otimes \ell_1,\ldots,a_i\otimes \ell_i)&:=\\
 &\kern-1cm:=(-1)^{i\sum_{j=1}^i|a_j|+\sum_{j=2}^{i} |a_j|\sum_{k=1}^{j-1} |\ell_k|}\,\times\\[-5pt]
  &\times(a_1 \cdots a_i)\otimes \mu_i(\ell_1,\ldots,\ell_i)
\end{aligned}
\end{equation}
\end{subequations}
for $i\geq 2$ and homogeneous $a_1,\ldots,a_i\in \sfA$ and $\ell_1,\ldots,\ell_i\in \sfL$, and these products extend to general elements by linearity. If, in addition, both $\sfA$ and $\sfL$ come with inner products, then $\!\!\hat{~\sfL}$ admits a natural inner product defined by
\begin{equation}
 \langle a_1\otimes\ell_1,a_2\otimes\ell_2\rangle:=(-1)^{|a_2| |\ell_1|}\langle a_1,a_2\rangle\langle\ell_1,\ell_2\rangle
\end{equation}
for homogeneous $a_1,a_2\in \sfA$ and $\ell_1,\ell_2\in \sfL$ and again extended to general elements by linearity. Detailed proofs on checking the higher Jacobi identities for the products $\hat\mu_i$ and the cyclicity of this inner product can be found in~\cite{Jurco:2018sby}.

The prime example of such a {\it tensor product $L_\infty$-algebra} is the tensor product of the de Rham complex $(\Omega^\bullet(X),\dd)$ on a $d$-dimensional manifold $X$ for $d\geq3$ with a finite-dimensional $L_\infty$-algebra $(\sfL,\mu_i)$ with $\sfL=\bigoplus_{k=-d+3}^0\sfL_k$. In this case we shall write $\Omega^\bullet(X,\sfL)$. If one assumes that $X$ is also compact, oriented, and without boundary, then there is a natural inner product on $\Omega^\bullet(X)$ given by $\langle\omega_1,\omega_2\rangle:=\int_X\omega_1\wedge\omega_2$. Provided $\sfL$ is cyclic then $\Omega^\bullet(X,\sfL)$ comes with a natural inner product by means of the above construction. We shall come back to this example later on when discussing higher Chern--Simons theory.

Morphisms between $L_\infty$-algebras, also known as {\it $L_\infty$-morphisms} generalise the notion of Lie algebra morphisms, and are most straightforwardly understood in the $Q$-manifold picture. In particular, an $L_\infty$-morphism is described by a degree 0 morphism $(f,f^\sharp):(X,Q)\to(X',Q')$ of $\IZ$-graded manifolds that preserves the homological vector fields in the sense that $Q\circ f^\sharp=f^\sharp \circ Q'$. In the $L_\infty$-picture this corresponds to a collection of multilinear totally graded antisymmetric maps $\phi_i:\sfL\times\cdots\times\sfL\to\sfL'$ of degree $1-i$ for two $L_\infty$-algebras $(\sfL,\mu_i)$ and $(\sfL',\mu_i')$ such that
\begin{subequations}\label{eq:LMor}
\begin{equation}
\begin{aligned}
   &\sum_{j+k=i}\sum_{\sigma\in {\rm Sh}(j;i)}~(-1)^{k}\chi(\sigma;\ell_1,\ldots,\ell_i)\times\\
   &\kern.5cm\times \phi_{k+1}(\mu_j(\ell_{\sigma(1)},\dots,\ell_{\sigma(j)}),\ell_{\sigma(j+1)},\dots ,\ell_{\sigma(i)})=\\
   &=\sum_{j=1}^i\frac{1}{j!} \sum_{k_1+\cdots+k_j=i}\sum_{\sigma\in{\rm Sh}(k_1,\ldots,k_{j-1};i)}\times\\
   &\kern.5cm\times\chi(\sigma;\ell_1,\ldots,\ell_i)\zeta(\sigma;\ell_1,\ldots,\ell_i)\,\times\\
   &\kern1cm\times \mu'_j\Big(\phi_{k_1}\big(\ell_{\sigma(1)},\ldots,\ell_{\sigma(k_1)}\big),\ldots,\\
   &\kern2.5cm \phi_{k_j}\big(\ell_{\sigma(k_1+\cdots+k_{j-1}+1)},\ldots,\ell_{\sigma(i)}\big)\Big)~,
\end{aligned}
\end{equation}
where $\chi(\sigma;\ell_1,\ldots,\ell_i)$ is the aforementioned Koszul sign and $\zeta(\sigma;\ell_1,\ldots,\ell_i)$ for a
$(k_1,\dots, k_{j-1};i)$-shuffle $\sigma$ is given by
\begin{equation}
\begin{aligned}
 &\zeta(\sigma;\ell_1,\ldots,\ell_i):=\\
 &\kern.5cm:=(-1)^{\sum_{1\leq m<n\leq j}k_mk_n+\sum_{m=1}^{j-1}k_m(j-m)}\times\\
 &\kern1cm\times(-1)^{\sum_{m=2}^j(1-k_m)\sum_{k=1}^{k_1+\cdots+k_{m-1}}|\ell_{\sigma(k)}|}~.
 \end{aligned}
\end{equation}
\end{subequations}

Since $\mu_1$ is a differential, we can consider the cohomology ring of an $L_\infty$-algebra $(\sfL,\mu_i)$, denoted by $H^\bullet_{\mu_1}(\sfL)$, and whenever the map $\phi_1$ for an $L_\infty$-morphism $(\sfL,\mu_i)\to(\sfL',\mu_i')$ induces an isomorphism $H^\bullet_{\mu_1}(\sfL)\cong H^\bullet_{\mu'_1}(\sfL')$, the $L_\infty$-morphism is called an {\it $L_\infty$-quasi-isomorphism}. Importantly, quasi-isomorphisms induce an equivalence relation on the space of all $L_\infty$-algebras. 

A differential graded Lie algebra, which is a $\IZ$-graded vector space equipped with graded Lie bracket and a differential that is a graded derivation with respect to the Lie bracket, is, evidently, an example of an $L_\infty$-algebra. Importantly, however, it can be shown~\cite{igor1995} that any $L_\infty$-algebra is $L_\infty$-quasi-isomorphic to a differential graded Lie algebra. This is known as the {\it strictification} of an $L_\infty$-algebra. Whilst this result is crucial for making general statement about $L_\infty$-algebras, in practical applications it often very difficult to construct the strictification $L_\infty$-quasi-isomorphism explicitly.

Besides this strictification theorem, there is another important theorem, known as the {\it minimal model theorem}~\cite{kadeishvili1982algebraic,Kajiura:0306332}, which says that any $L_\infty$-algebra $(\sfL,\mu_i)$ is quasi-isomorphic to an $L_\infty$-algebra $(\sfL',\mu'_i)$ with $\mu'_1=0$. An $L_\infty$-algebra with $\mu_1=0$ is known as a {\it minimal model}. Essentially, $\sfL'$ is, unique up to $L_\infty$-isomorphism, the cohomology ring $H^\bullet_{\mu_1}(\sfL)$ of $(\sfL,\mu_i)$ and the $L_\infty$-quasi-isomorphism determined by the maps $\phi_i:\sfL'\times\cdots\times\sfL'\to\sfL$ and products $\mu'_i$ are constructed recursively as~\cite{Kajiura:0306332}
\begin{subequations}
\begin{equation}
\begin{aligned}
 \phi_1(\ell'_1)&:=e(\ell'_1)~,\\
  \phi_2(\ell'_1,\ell'_2)&:=- h(\mu_2(\phi_1(\ell'_1),\phi_1(\ell'_2)))~,\\
  &~~\phantom{:}\vdots\\
  \phi_i(\ell'_1,\ldots,\ell'_i)&:=-\sum_{j=2}^i\frac{1}{j!} \sum_{k_1+\cdots+k_j=i}\sum_{\sigma\in{\rm Sh}(k_1,\ldots,k_{j-1};i)}\times\\
  &\kern-1cm\times\chi(\sigma;\ell'_1,\ldots,\ell'_i)\zeta(\sigma;\ell'_1,\ldots,\ell'_i)\,\times\\
   &\kern-1cm\times h\left\{\mu_j\Big(\phi_{k_1}\big(\ell'_{\sigma(1)},\ldots,\ell'_{\sigma(k_1)}\big),\ldots,\right.\\
   & \left.\phi_{k_j}\big(\ell'_{\sigma(k_1+\cdots+k_{j-1}+1)},\ldots,\ell'_{\sigma(i)}\big)\Big)\right\}
\end{aligned}
\end{equation}
and
\begin{equation}
\begin{aligned}
 \mu'_1(\ell'_1)&:=0~,\\
 \mu'_2(\ell'_1,\ell'_2)&:=p(\mu_2(\phi_1(\ell'_1),\phi_1(\ell'_2)))~,\\
  &~~\phantom{:}\vdots\\
   \mu'_i(\ell'_1,\ldots,\ell'_i)&:=\sum_{j=2}^i\frac{1}{j!} \sum_{k_1+\cdots+k_j=i}\sum_{\sigma\in{\rm Sh}(k_1,\ldots,k_{j-1};i)}\times\\
   &\kern-1cm\times\chi(\sigma;\ell'_1,\ldots,\ell'_i)\zeta(\sigma;\ell'_1,\ldots,\ell'_i)\,\times\\
   &\kern-1cm\times p\left\{\mu_j\Big(\phi_{k_1}\big(\ell'_{\sigma(1)},\ldots,\ell'_{\sigma(k_1)}\big),\ldots,\right.\\
   &\left.\phi_{k_j}\big(\ell'_{\sigma(k_1+\cdots+k_{j-1}+1)},\ldots,\ell'_{\sigma(i)}\big)\Big)\right\},
\end{aligned}
\end{equation}
where $\ell'_1,\ldots,\ell'_i\in\sfL'$. Here, $\chi(\sigma;\ell'_1,\ldots,\ell'_i)$ is the Koszul sign and $\zeta(\sigma;\ell'_1,\ldots,\ell'_i)$ the sign factor introduced above, and $h$ and $e$ are maps appearing in
\begin{equation}
 \kern-6pt\myxymatrix{\ar@(dl,ul)[]^h \sfL\ar@<+2pt>@{->>}[rr]^{\kern-20pt p} & & ~~H^\bullet_{\mu_1}(\sfL) \ar@<+2pt>@{^(->}[ll]^{\kern-20pt e}}
\end{equation}
\end{subequations}
with $p\circ e=1$ and $h$ is a {\it contracting homotopy}. The latter means that $h$ is a collection of degree $-1$ morphisms $h_k:\sfL_k\rightarrow \sfL_{k-1}$ that obey $\mu_1= \mu_1\circ h \circ \mu_1$. It then follows that we can introduce the three projectors
\begin{equation}
\begin{aligned}
 &P_1:=e\circ p~,~~~P_2:=h\circ \mu_1~,~~~P_3:=\mu_1\circ h~,\\
 &P_i\circ P_j=\delta_{ij}P_i~,~~~\id=P_1+P_2+P_3
 \end{aligned}
\end{equation}
implying the decomposition
\begin{equation}
  \sfL\cong H^\bullet_{\mu_1}(\sfL) \oplus \im(h\circ \mu_1)\oplus \im(\mu_1\circ h)~.
\end{equation}
This is known as the {\it abstract Hodge--Kodaira decomposition}, see e.g.~\cite{Kajiura:0306332}.

\section{Quasi-groups}

Having discussed $L_\infty$-algebras as higher generalisations of Lie algebras, we now face the question about their finite counter parts. In particular, Lie algebras integrate to Lie groups and, vice versa, Lie groups differentiate to Lie algebras. It turns out that this question in the context of $L_\infty$-algebras is rather involved. Eventually, the finite counter part of an $L_\infty$-algebra is equivalent to a quasi-group \cite{Severa:2006aa,Henriques:2006aa,Severa:1506.04898}. To define the latter, we shall need the machinery of simplicial geometry which we briefly recap for the reader's convenience. For more details, see e.g.~\cite{Friedman:0809.4221} or the text books \cite{0387984038,0226511812,Goerss:1999aa}.

\subsection{Simplicial manifolds}

Let us start by introducing the {\it simplex category} $\Delta$. This is the category which has totally ordered sets $[p]:=\{0,1,\ldots, p\}$ for $p=0,1,2,\ldots$ as objects and order-preser\-ving maps $[p]\to[p']$ as morphisms. The latter are generated by the {\it coface maps} $\phi^p_i$ and {\it codegeneracy maps} $\delta^p_i$ both of which are given by
\begin{equation}\label{eq:CoFandCoD}
\begin{minipage}{11cm}
\tikzset{->-/.style={decoration={markings,mark=at position #1 with {\arrow{>}}},postaction={decorate}}}
\begin{tikzpicture}[scale=.8,every node/.style={scale=1}]
   \draw (1,1) node {$\phi^p_i:[p-1]\ \to\ [p]$};
   \draw (-.3,0) node {$0$};
   \draw (-.3,-.5) node {$1$};
   \draw (0,-.9) node {$\vdots$};
   \draw (2,-.9) node {$\vdots$};
   \draw (-.6,-1.5) node {$i-1$};
   \draw (-.3,-2) node {$i$};
   \draw (0,-2.4) node {$\vdots$};
   \draw (2,-2.9) node {$\vdots$};
   \draw (-.6,-3) node {$p-1$};
   \draw (2.3,0) node {$0$};
   \draw (2.3,-.5) node {$1$};
   \draw (2.6,-1.5) node {$i-1$};
   \draw (2.28,-2) node {$i$};
   \draw (2.6,-2.5) node {$i+1$};
   \draw (2.28,-3.5) node {$p$};
   \draw[->-=.5] (0,0) -- (2,0);   
   \draw[->-=.5] (0,-.5) -- (2,-.5);   
   \draw[->-=.5] (0,-1.5) -- (2,-1.5);   
   \draw[->-=.5] (0,-2) -- (2,-2.5);   
   \draw[->-=.5] (0,-3) -- (2,-3.5);   
   \filldraw [black] (0,0) circle (1.5pt);
   \filldraw [black] (0,-.5) circle (1.5pt);
   \filldraw [black] (0,-1.5) circle (1.5pt);
   \filldraw [black] (0,-2) circle (1.5pt);
   \filldraw [black] (0,-3) circle (1.5pt);
   \filldraw [black] (2,0) circle (1.5pt);
   \filldraw [black] (2,-.5) circle (1.5pt);
   \filldraw [black] (2,-1.5) circle (1.5pt);
   \filldraw [black] (2,-2) circle (1.5pt);
   \filldraw [black] (2,-2.5) circle (1.5pt);
   \filldraw [black] (2,-3.5) circle (1.5pt);
   \draw (7,1) node {$\delta^p_i:[p+1]\ \to\ [p]$};
   \draw (5.7,0) node {$0$};
   \draw (5.7,-.5) node {$1$};
   \draw (5.7,-1.5) node {$i$};
   \draw (5.4,-2) node {$i+1$};
   \draw (5.4,-2.5) node {$i+2$};
   \draw (5.4,-3.5) node {$p+1$};
   \draw (8.3,0) node {$0$};
   \draw (8.3,-.5) node {$1$};
   \draw (8.25,-1.5) node {$i$};
   \draw (8.6,-2) node {$i+1$};
   \draw (8.3,-3) node {$p$};
   \draw (6,-.9) node {$\vdots$};
   \draw (6,-2.9) node {$\vdots$};
   \draw (8,-.9) node {$\vdots$};
   \draw (8,-2.4) node {$\vdots$};
   \draw[->-=.5] (6,0) -- (8,0);   
   \draw[->-=.5] (6,-.5) -- (8,-.5);   
   \draw[->-=.5] (6,-1.5) -- (8,-1.5);   
   \draw[->-=.5] (6,-2) -- (8,-1.5);   
   \draw[->-=.5] (6,-2.5) -- (8,-2);  
   \draw[->-=.5] (6,-3.5) -- (8,-3);   
   \filldraw [black] (6,0) circle (1.5pt);
   \filldraw [black] (6,-.5) circle (1.5pt);
   \filldraw [black] (6,-1.5) circle (1.5pt);
   \filldraw [black] (6,-2) circle (1.5pt);
   \filldraw [black] (6,-2.5) circle (1.5pt);
   \filldraw [black] (6,-3.5) circle (1.5pt);
   \filldraw [black] (8,0) circle (1.5pt);
   \filldraw [black] (8,-.5) circle (1.5pt);
   \filldraw [black] (8,-1.5) circle (1.5pt);
   \filldraw [black] (8,-2) circle (1.5pt);
   \filldraw [black] (8,-3) circle (1.5pt);
  \end{tikzpicture}
  \end{minipage}
\end{equation}
Indeed any order-preserving map $\phi:[p]\to[p']$ can be decomposed as
\begin{equation}
 \phi=\phi_{i_m}\circ\cdots\circ\phi_{i_1}\circ\delta_{j_1}\circ\cdots\circ\delta_{j_n}
\end{equation}
with $p+m-n=p'$, $0\leq i_1<\cdots<i_m\leq p'$, and $0\leq j_1<\cdots<j_n< p$. In addition, if we let $\sfTop$ be the category of topological spaces, then the objects in the simplex category $\Delta$ have a \emph{geometric realisation} in terms of the standard topological $p$-simplices,
\begin{equation}
 |\Delta^p|:=\left\{(t_0,\ldots,t_p)\in\IR^{p+1}\,|\, \sum_{i=0}^p t_i=1~\mbox{and}~ t_i\geq 0\right\},
\end{equation}
by means of the functor $\Delta\to\sfTop$ defined by $[p]\mapsto|\Delta^p|$ and 
\begin{equation}
\begin{aligned}
  &\Big([p]\ \overset{\phi}{\longrightarrow}\ [p']\Big)\mapsto\\
  &\kern.5cm\mapsto\begin{pmatrix}|\Delta^p|& \overset{}{\longrightarrow}& |\Delta^{p'}|\\ (t_0,\ldots,t_p)& \mapsto& \big(\sum_{\phi(i)=0}t_i,\ldots,\sum_{\phi(i)=p'}t_i\big)\end{pmatrix}.
  \end{aligned}
\end{equation}
Thus, the coface map $\phi^p_i$ induces the injection $|\Delta^p|\hookrightarrow |\Delta^{p+1}|$ given by $(t_0,\ldots,t_p)\mapsto (t_0,\ldots,t_{i-1},0, t_{i},\ldots, t_p)$ and sending $|\Delta^p|$ to the $i$-th face of $|\Delta^{p+1}|$. Likewise, the codegeneracy map $\delta^p_i$ induces the projection $|\Delta^p|\rightarrow |\Delta^{p-1}|$ by $(t_p,\ldots,t_0)\mapsto (t_0,\ldots,t_{i}+t_{i+1},\ldots, t_p)$ sending $|\Delta^p|$ to $|\Delta^{p-1}|$ by collapsing together the vertices $i$ and $i+1$.

With this in mind, let $\sfSet$ be the category of sets. A {\it simplicial set} $\scrX$ is simply a $\sfSet$-valued presheaf on $\Delta$, that is, is a functor $\scrX:\Delta^{\rm op}\to\sfSet$ where the superscript `op' refers to the opposite category in which the objects are the same but the morphisms reversed. We could replace $\sfSet$ by the category of groups $\sfGrp$ or the category of (Frech\'et) manifolds $\sfMfd$ to obtain simplicial groups or simplicial manifolds, respectively. Explicitly, this definition means that $\scrX$ is a collection of sets $\scrX_p:=\scrX([p])$ called the {\it simplicial $p$-simplices} and maps $\sff^p_i:=\scrX(\phi^p_i):\scrX_p\rightarrow \scrX_{p-1}$ called the {\it face maps} and $\sfd^p_i:=\scrX(\delta^p_i):\scrX_p\to\scrX_{p+1}$ called the degeneracy maps subject to the \emph{simplicial identities}
\begin{equation}
 \begin{aligned}
  \sff_i\circ\sff_j&=\sff_{j-1}\circ\sff_i\efor i\ <\ j~,\\
  \sfd_i\circ\sfd_j&=\sfd_{j+1}\circ\sfd_i\efor i\ \leq\ j~,\\
  \sff_i\circ\sfd_j&=\sfd_{j-1}\circ\sff_{i}\efor i\ <\ j~,\\
  \sff_i\circ\sfd_j&=\sfd_j\circ\sff_{i-1}\efor i\ >\ j+1~,\\
  \sff_i\circ\sfd_i&=\id=\sff_{i+1}\circ\sfd_i~.
 \end{aligned}
\end{equation}
These identities straightforwardly follow from similar identities for the coface and codegeneracy maps. In the following, we shall depict simplicial sets by writing arrows for the face maps, that is,
\begin{equation}
 \left\{ \cdots\quadarrow\scrX_2\triplearrow\scrX_1\doublearrow\scrX_0\right\}.
\end{equation}

We define morphisms of simplicial sets, also called {\it simplicial maps}, to be the natural transformation between the functors defining the simplicial sets as pre\-sheaves. Put differently, a simplicial map $g:\scrX\to\scrX'$ between two simplicial sets is a collection of maps $g^p:\scrX_p\to\scrX'_p$ that commute with the face and degeneracy maps on $\scrX$ and $\scrX'$. Simplicial sets together with simplicial maps for the category of simplicial sets $\sfsSet$. More succinctly, $\sfsSet$ is the functor category $\sfFun(\Delta^{\rm op},\sfSet)$.

The prime examples of a simplicial set is the {\it standard simplicial $p$-simplex} $\Delta^p$ which is the simplicial set $\sfhom_\Delta(-,[p]):\Delta^{\rm op}\to\sfSet$. This simplicial set has a unique non-degenerate simplicial $p$-simplex. By virtue of the Yoneda lemma, any simplicial map $\Delta^p\to\Delta^{p'}$ corresponds bijectively to a morphism $[p]\to[p']$ in the simplex category $\Delta$. Moreover, the Yoneda lemma also implies the bijection 
\begin{equation}\label{eq:SimSim}
\scrX_p\cong\sfhom_{\sfsSet}(\Delta^p,\scrX)
\end{equation}
that for any simplical set $\scrX$.

Given any two simplicial sets $\scrX$ and $\scrX'$, we may form their {\it product} $\scrX\times \scrX'$ by defining it to be the simplicial set with simplicial $p$-simplices $(\scrX\times \scrX')_p:=\scrX_p\times \scrX'_p$ together with the face and degeneracy maps acting as $\sff_i^{\scrX\times \scrX'}(x_p,x'_p):=(\sff_i^\scrX x_p,\sff_i^{\scrX'} x'_p)$ and $\sfd_i^{\scrX\times \scrX'}(x_p,x'_p):=(\sfd_i^\scrX x_p,\sfd_i^{\scrX'} x'_p)$ for all $(x_p,x'_p)\in (\scrX\times \scrX')_p$. This makes $\sfsSet$ into a (strict) monoidal category.

Furthermore, for any two simplicial sets $\scrX$ and $\scrX'$ we define the simplicial set $\inthom(\scrX,\scrX')$, called the {\it internal hom}, by letting $\inthom_p(\scrX,\scrX'):=\sfhom_\sfsSet(\Delta^p\times \scrX,\scrX')$ be its simplicial $p$-simplices and its  face and degeneracy maps are given by
\begin{equation}
\begin{aligned}
&\sff^p_i\,:\, \Big(\Delta^p\times \scrX\  \overset{f}{\longrightarrow}\ \scrX'\Big)\mapsto\\
&\kern.5cm\mapsto \Big(\Delta^{p-1}\times \scrX\ \overset{\phi^p_i\times\id_\scrX}{\longrightarrow}\ \Delta^p\times \scrX\  \overset{f}{\longrightarrow}\ \scrX'\Big)~,\\
  &\sfd^p_i\,:\,\Big(\Delta^p\times \scrX\  \overset{f}{\longrightarrow}\ \scrX'\Big)\mapsto\\
  &\kern.5cm\mapsto\Big(\Delta^{p+1}\times \scrX\ \overset{\delta^p_i\times\id_\scrX}{\longrightarrow}\ \Delta^p\times \scrX\  \overset{f}{\longrightarrow}\ \scrX'\Big)~.
  \end{aligned}
\end{equation}
Evidently, the simplicial 0-simplices $\inthom_0(\scrX,\scrX')$ are the simplicial maps between $\scrX$ and $\scrX'$. By virtue of the Yoneda lemma, it follows that
\begin{equation}
\sfhom_\sfsSet(\Delta^p\times \scrX,\scrX')\cong\sfhom_\sfsSet(\Delta^p,\inthom(\scrX,\scrX'))~,
\end{equation}
and this can be generalised further to
\begin{equation}\label{eq:IsoIntHom}
\sfhom_\sfsSet(\scrX\times\scrX',\scrX'')\cong\sfhom_\sfsSet(\scrX,\inthom(\scrX',\scrX''))
\end{equation}
for any three simplicial sets $\scrX$, $\scrX'$, and $\scrX''$. 

We are now ready to introduce simplicial homotopies. A {\it simplicial homotopy} between two simplicial maps $g,\tilde g:\scrX\to\scrX'$  for two simplicial sets $\scrX$ and $\scrX'$ is an element $h\in\inthom_1(\scrX,\scrX')=\sfhom_\sfsSet(\Delta^1\times\scrX,\scrX')$ that renders the diagram
 \begin{equation}
    \xymatrixcolsep{5pc}
    \myxymatrix{
    \Delta^0\times \scrX\cong\scrX\ar@{->}[rd]^{g} \ar@{->}[d]_{\phi_1^1\times\id_\scrX} & \\
    \Delta^1\times \scrX\ar@{->}[r]^{h} & \scrX'\\
    \Delta^0\times \scrX\cong\scrX\ar@{->}[ru]^{\tilde g} \ar@{->}[u]^{\phi_0^1\times\id_\scrX} & 
    }
 \end{equation}
commutative. Equivalently, using~\eqref{eq:IsoIntHom}, a simplicial map $h\in\sfhom_\sfsSet(\scrX,\inthom(\Delta^1,\scrX'))$, which is a collection of maps $h^p=(h^p_i):\scrX_p\to \inthom_p(\Delta^1,\scrX')$ with $h^p_i:\scrX_p\to\scrX'_{p+1}$ for $i=0,\ldots,p$, is a simplicial homotopy between the simplicial maps $g^p:=\sff^{p+1}_0\circ h^p_0:\scrX_p\to\scrX'_p$ and $\tilde g_p:=\sff^{p+1}_{p+1}\circ h^p_p:\scrX_p\to\scrX'_p$. In this spirit, {\it higher simplicial homotopies} will be elements of $\inthom_k(\scrX,\scrX')\cong\sfhom_\sfsSet(\scrX, \inthom(\Delta^k,\scrX'))$ for $k\geq2$.\footnote{Since $\inthom(\Delta^0,\scrX)\cong\scrX$, simplicial maps, simplicial homotopies, and all the higher simplicial homotopies are given by $\sfhom_\sfsSet(\scrX, \inthom(\Delta^k,\scrX'))$ for $k\geq0$.}

In~\eqref{eq:SimSim} we have seen how the simplicial simplices of a simplicial set $\scrX$ can be understood in terms of simplicial maps from the standard simplicial simplex $\Delta^p$ to $\scrX$. For each $i$, we may define the {\it $(p,i)$-horn $\Lambda^p_i$} of $\Delta^p$ to be the simplicial subset of $\Delta^p$ that is generated by the union of all faces of $\Delta^p$ except for the $i$-th one, and, more generally, the $(p,i)$-horns of a simplicial set $\scrX$ are the elements of $\sfhom_{\sfsSet}(\Lambda^p_i,\scrX)$. Evidently, since all the horns $\Lambda^p_i$ of $\Delta^p$ arise by removing the unique non-degenerate simplicial $p$-simplex from $\Delta^p$ and the $i$-th non-degenerate simplicial $(p-1)$-simplex, they can be completed again to simplicial simplices. However, the horns $\sfhom_{\sfsSet}(\Lambda^p_i,\scrX)$ of a general simplicial set $\scrX$ may not always be completed to simplicial simplices $\sfhom_{\sfsSet}(\Delta^p,\scrX)$. Whenever this can be done, that is, whenever there is a simplicial map $\tilde\delta:\Delta^p\to\scrX$ for any horn $\lambda:\Lambda^p_i\to\scrX$ such that
 \begin{equation}
    \myxymatrix{
    \Lambda^p_i\ar@{->}[r]^{\lambda} \ar@{^{(}->}[d] & \scrX \\
    \Delta^p \ar@{->}[ru]^{\tilde\delta} & 
    }
 \end{equation}
 is commutative, we call $\scrX$ a {\it Kan simplicial set}. Put differently, the natural restriction mappings
\begin{equation}
\sfhom_\sfsSet(\Delta^p,\scrX)\to\sfhom_\sfsSet(\Lambda^p_i,\scrX)
\end{equation}
are surjective for all $p\geq1$ and $0\leq i\leq p$. For {\it Kan simplicial manifolds}, we replace the category of sets by the category of (Fr\'echet) manifolds, and we also require the above restrictions to be submersions. Notice that whenever $\scrX'$ is Kan, so is the internal hom $\inthom(\scrX,\scrX')$.

An important example of a Kan simplicial manifold is the nerve of the \v Cech groupoid: let  $\phi\,:\,Y\to X$ be a surjective submersion between two manifolds $Y$ and $X$ and denote the fibre product of $Y$ with itself over $X$ by $Y\times_X Y:=\{(y_1,y_2)\in Y\times Y\,|\,\phi(y_1)=\phi(y_2)\}$. The {\it \v Cech groupoid} $\check{\calC}(Y\to X)$ of $f$ is the groupoid $Y\times_X Y\doublearrow Y$ with pairs $(y_1,y_2)$ for $y_1,y_2\in Y$ satisfying $\phi(y_1)=\phi(y_2)$ as its morphisms. It has the source, target, composition, and identity maps given by  $\sfs(y_1,y_2):=y_2$, $\sft(y_1,y_2):=y_1$, $\id_y:=(y,y)$, and $(y_1,y_2)\circ(y_2,y_3):=(y_1,y_3)$. The {\it nerve} of the \v Cech groupoid, also known as the {\it \v Cech nerve}, is the simplicial set
\begin{subequations}\label{eq:CechNerve}
\begin{equation}
\begin{aligned}
&N(\check{\calC}(Y\to X)):=\\
&\kern.5cm:=\left\{ \cdots \quadarrow Y\times_X Y\times_X Y\triplearrow Y\times_X Y\doublearrow Y\right\}
\end{aligned}
\end{equation}
with face and degeneracy maps defined as
\begin{equation}
\begin{aligned}
 \sff^p_i(y_0,\ldots,y_{p})&:=(y_0,\ldots,y_{i-1},y_{i+1},\ldots,y_{p})~,\\
  \sfd^p_i(y_0,\ldots,y_{p})&:=(y_0,\ldots,y_{i-1},y_i,y_{i},\ldots,y_{p})~.
 \end{aligned}
\end{equation}
\end{subequations}
It can be shown that this is a Kan simplicial manifold.

\subsection{Quasi-groups and $L_\infty$-algebras}\label{sec:GroupoidsLinfty}

Importantly, whilst in general simplicial homotopy does not induce an equivalence relation on $\sfhom_{\sfsSet}(\scrX,\scrX')$ it always does when $\scrX'$ is a Kan simplicial set. Amongst other things, this fact will be essential below when introducing higher principal bundles.

Kan simplicial sets are also known as {\it quasi-groupoids} and Kan simplicial manifolds as {\it Lie quasi-groupoids}, respectively. Furthermore, if there is only one single simplicial 0-simplex, a Kan simplicial set (manifold) is called a {\it reduced (Lie) quasi-groupoid}. We shall follow the delooping hypothesis and identify reduced (Lie) quasi-groupoids with {\it (Lie) quasi-groups}. Importantly, the categories of (Lie) quasi-groups and simplicial (Lie) groups are equivalent due to a classical result of Quillen's \cite{quillen1969}. In addition, whenever all the $(p,i)$-horns for a (Lie) quasi-group can be filled uniquely for all $p>n$, we shall speak of a {\it (Lie) $n$-quasi-group}. 

In Section \ref{sec:Qmfd}, we have introduced the notion of a $\IZ$-graded manifold. Using the forgetful functor, we may map $\IZ$-graded manifolds to $\IZ_2$-graded manifolds which are also known as {\it supermanifolds}. We let $\sfSMfd$ be the category of (Frech\'et) supermanifolds. Moreover, denote by $\sfSurSub$ the category of surjective submersions $Y\to X$ between supermanifolds $Y$ and $X$ as its objects and maps as its morphisms such that 
\begin{equation}
    \myxymatrix{
    Y_1\ar@{->}[r] \ar@{->}[d] & Y_2\ar@{->}[d]\\
    X_1 \ar@{->}[r] & X_2
    }
\end{equation}
are commutative for surjective submersions $Y_{1,2}\to X_{1,2}$. As before, we set $\sfsSMfd:=\sfFun(\Delta^{\rm op},\sfSMfd)$ and call it the category of {\it simplicial supermanifolds}. 

Since the nerve $N$ of the \v Cech groupoid of an object in $\sfSurSub$ is an object in $\sfsSMfd$, any object $\scrX\in\sfsSMfd$ can be used to define a $\sfSet$-valued presheaf $\sfhom_\sfsSMfd(N(-),\scrX):\sfSurSub^{\rm op}\to\sfSet$ on $\sfSurSub$. We are now interested in the linearisations of this presheaf, which we shall call the {\it $k$-jets} of $\scrX$ in spirit of an analogous construction in ordinary differential geometry. Specifically, let us consider the subcategory $\sfSurSub_k$ of $\sfSurSub$ defined to be the category whose objects are surjective submersions of the form $X\times\IR^{0|k}\to X$. We have the identification
\begin{equation}
\begin{aligned}
 &\sfhom_{\sfSurSub_k}(X_1\times\IR^{0|k}\to X_1,X_2\times\IR^{0|k}\to X_2)\cong\ \\
  &\kern.5cm\cong\sfhom_{\sfsSMfd}(X_1,X_2)\times \sfhom_{\sfSMfd}(X_1\times \IR^{0|k},\IR^{0|k})~.
\end{aligned}
\end{equation}
Evidently, this implies that a presheaf on $\sfSurSub_k$ is equivalent to a presheaf on $\sfSMfd$ together with an action of  $\inthom(\IR^{0|k},\IR^{0|k})$. We shall denote this by $\sfSMfd_k$. For instance, $\sfSMfd_1$ is the category of $Q$-supermanifolds since the action of $\inthom(\IR^{0|1},\IR^{0|1})$ corresponds to the action of the vector field $Q$. Following \v Severa \cite{Severa:2006aa}, for any presheaf on $\sfSurSub$, we may consider its restriction to $\sfSurSub_k$ to obtain a presheaf on $\sfSMfd_k$ the latter of which we call the {\it $k$-jet} of the presheaf on $\sfSurSub$. In addition, the {\it $k$-jet} of a simplicial supermanifold $\scrX$ is the $k$-jet of the presheaf $\sfhom_\sfsSMfd(N(-),\scrX)$.

It turns out that the 1-jet of a Lie quasi-group is an $L_\infty$-algebra \cite{Severa:2006aa}; see also \cite{Jurco:2016qwv} for a constructive proof. In particular, letting
\begin{equation}
 \scrG:=\left\{ \cdots\quadarrow\scrG_2\triplearrow\scrG_1\doublearrow * \right\}
\end{equation}
be a Lie quasi-group with face maps $\sff^p_i$ and degeneracy maps $\sfd^p_i$, the 1-jet of $\scrG$ is parametrised as~\cite{Jurco:2016qwv}
\begin{equation}
 \sfL[1]=\bigoplus_{k\leq0}\sfL_k[1]\ewith \sfL_k[1]:=\bigcap_{i=0}^{-k}\ker\big(\sff^{1-k}_{i\,*}\big)[1-k]~,
\end{equation}
where $\sff^{p}_{i\,*}$ denotes the linearisation of $\sff^p_i$. Furthermore, $\mu_1|_{\sfL_k[1]}=\sff^{1-k}_{1-k\,*}$ and the $\mu_i$ for $i>1$ are given in terms of $j$-th order derivatives of the face maps with $j\leq i$.\footnote{Note that $\mu_1|_{\sfL_0[1]}=\sff^{1}_{1\,*}=0$ as $\scrG$ has only one simplicial 0-simplex.}

The converse is also true though this is a much more involved problem due to topological questions: every $L_\infty$-algebra integrates to a Lie quasi-group. See \cite{Henriques:2006aa,Severa:1506.04898} for details.

\section{Higher principal bundles}

Let us now discuss how principal bundles with quasi-groups as their structure groups are formulated. The following constructions have a long history, and we refer to e.g.~\cite{Murray:9407015,Aschieri:2003mw,Baez:2004in,Bartels:2004aa,Breen:math0106083,Baez:0511710,Wockel:2008aa,Schreiber:2008aa,Saemann:2012uq,Nikolaus:1207ab,nikolaus1207,Saemann:2013pca,Wang:2013dwa,Jurco:2014mva,Jurco:2016qwv} for details.

\subsection{Principal $\scrG$-bundles}

Let $\sfG$ be a Lie group and consider its delooping $\sfB\sfG$ which is the Lie groupoid $\sfG\doublearrow *$ for which the source and target maps are trivial, $\id_*=\unit_\sfG$, and the composition is group multiplication in $\sfG$. Consider its nerve 
\begin{equation}
N(\sfB\sfG):=\left\{ \cdots \quadarrow \sfG\times\sfG\times\sfG\triplearrow \sfG\times\sfG\doublearrow\sfG\right\}
\end{equation}
with the obvious face and degeneracy maps. 

Furthermore, recall the \v Cech nerve~\eqref{eq:CechNerve} for a surjective submersion $\dot\bigcup_{a\in A}U_a\to X$ given by an open cover $\{U_a\}_{a\in A}$ of $X$. With these ingredients, a principal $\sfG$-bundle is a simplicial map $g:N(\check{\calC}(\dot\bigcup_{a\in A}U_a\to X))\to N(\sfB\sfG)$. Indeed, $g$ is a collection of maps \linebreak $g^p:N_p(\check{\calC}(\dot\bigcup_{a\in A}U_a\to X))\to N_p(\sfB\sfG)$ explicitly given by
\begin{equation}
\begin{aligned}
g_a(x)&:=g^0(x,a)=*~,\\
g_{ab}(x)&:=g^1(x,a,b) \in \sfG~,\\
 g_{abc}(x)&:= g^2(x,a,b,c) =\big(g_{abc}^1(x),g_{abc}^2(x)\big) \in \sfG\times \sfG~.
 \end{aligned}
\end{equation}
Being simplicial, the $g^p$ commute with the face and degeneracy maps so that
\begin{equation}
\begin{aligned}
g_{abc}^1(x) &= g_{ab}(x)~,\\
g^1_{abc}(x) g^2_{abc}(x) &= g_{ac}(x)~,\\
g_{abc}^2(x) &= g_{bc}(x)~,
\end{aligned}
\end{equation}
that is, we obtain the standard cocycle conditions in terms of the transition functions $g_{ab}:U_a\cap U_b\to\sfG$. 

Moreover, it is an easy exercise to check that a simplicial homotopy $h:\Delta^1\times N(\check{\calC}(\dot\bigcup_{a\in A}U_a\to X))\to N(\sfB\sfG)$ between two principal $\sfG$-bundles $g,\tilde g:N(\check{\calC}(\dot\bigcup_{a\in A}U_a\to X))\to N(\sfB\sfG)$ amounts to a collection of maps $h_a:U_a\to\sfG$ with
\begin{equation}\label{eq:Cobdy1}
 g_{ab}(x)h_b(x)=h_a(x)\tilde g_{ab}(x)~,
\end{equation}
that is, the standard coboundary conditions.

Generally, for any Lie quasi-group $\scrG$, we define a {\it principal $\scrG$-bundle} over a manifold $X$ subordinate to an open cover $\dot\bigcup_{a\in A}U_a\to X$ to be a simplicial map $g:N(\check{\calC}(\dot\bigcup_{a\in A}U_a\to X))\to\scrG$ \cite{Nikolaus:1207ab,nikolaus1207}. Two such bundles are said to be {\it equivalent}, whenever there is a simplicial homotopy between the defining simplicial maps. It should be emphasised that this notion of equivalence is well-defined since $\scrG$ is Kan.

\subsection{Higher non-Abelian Deligne cohomology}

Besides principal bundles, we shall also need connective structures to discuss gauge theory. Recall that a connection or {\it connective structure} on a principal $\sfG$-bundle on a manifold $X$ subordinate to an open cover $\dot\bigcup_{a\in A}U_a\to X$ is a collection of $\frg$-valued differential 1-forms $\{A_a\in\Omega^1(U_a,\frg)\}$, with $\frg$ being the Lie algebra of $\sfG$, which obey 
\begin{equation}
A_b(x)=g^{-1}_{ab}(x)A_a(x)g_{ab}(x)+g^{-1}_{ab}(x)\dd g_{ab}(x)
\end{equation}
on non-empty intersections $U_a\cap U_b=\emptyset$. Here, the $g_{ab}$ are the transition functions of the principal $\sfG$-bundle. 

In addition, the coboundary transformations~\eqref{eq:Cobdy1} yield the transformations
\begin{equation}\label{eq:Cobdy2}
\tilde A_a(x)=h^{-1}_{a}(x)A_a(x)h_{a}(x)+h^{-1}_{a}(x)\dd h_{a}(x)~.
\end{equation}
This allows use to introduce the {\it Deligne cocycle} $\{A_a,g_{ab}\}$ which defines a principal $\sfG$-bundle with connection, and two such cocycles are called equivalent if there is a coboundary transformation of the form~\eqref{eq:Cobdy1} and~\eqref{eq:Cobdy2}.

It is clear now how this generalises to higher principal bundles. Concretely, let $\scrG$ be a Lie $n$-quasi-group and $(\sfL,\mu_i)$ with $\sfL=\bigoplus_{k=-n+1}^0\sfL_k$ be the associated $n$-term $L_\infty$-algebra obtained by computing the 1-jet of $\scrG$ (see Section \ref{sec:GroupoidsLinfty}). As before, let $X$ be a manifold with an open cover $\dot\bigcup_{a\in A}U_a\to X$. The transition functions 
\begin{subequations}\label{eq:HTrans}
\begin{equation}
g_{a_0\cdots a_k}:U_{a_0}\cap\ldots\cap U_{a_k}\to\scrG_k~,
\end{equation}
for $k=1,\ldots,n$, which are encoded in a simplicial map $g:N(\check{\calC}(\dot\bigcup_{a\in A}U_a\to X))\to\scrG$ defining a principal $\scrG$-bundle, are supplemented, when $n\geq2$, by differential-form-valued transition functions
\begin{equation}
\lambda_{a_0\cdots a_k}\in\bigoplus_{i+j=1-k}\Omega^i(U_{a_0}\cap\ldots\cap U_{a_k}) \otimes\sfL_{j}~,
\end{equation}
\end{subequations}
for $k=1,\ldots,n-1$. A {\it connective structure} on the principal $\scrG$-bundle is given by a set of local $L_\infty$-valued differential forms
\begin{equation}\label{eq:HCon}
 A_a\in \bigoplus_{i+j=1}\Omega^i(U_a)\otimes \sfL_{j}~,
\end{equation}
and~\eqref{eq:HCon} together with~\eqref{eq:HTrans} forms what is known as a {\it higher Deligne cocycle}.  Rather than listing the somewhat involved cocycle conditions and coboundary conditions for such a cocyle in full generality, let us instead exemplify our discussion with the example of a strict Lie 2-quasi-group~\cite{Murray:9407015,Aschieri:2003mw,Baez:2004in,Bartels:2004aa,Breen:math0106083,Baez:0511710,Wockel:2008aa,Schreiber:2008aa,Saemann:2012uq}. See~\cite{Nikolaus:1207ab,nikolaus1207,Saemann:2013pca,Wang:2013dwa,Jurco:2014mva,Jurco:2016qwv} for details for the general case. 

A strict Lie 2-quasi-group can equivalently be described by a Lie crossed module and the corresponding strict 2-term $L_\infty$-algebra by a differential crossed module. Specifically, a {\it Lie crossed module} is a pair of Lie groups $(\sfG,\sfH)$ together with an automorphism action $\acton$ of $\sfG$ on $\sfH$ and a group homomorphism $\sft:\sfH\rightarrow \sfG$ such that the homomorphism  $\sft$ is equivariant with respect to conjugation, $\sft(g\acton h)=g \sft(h) g^{-1}$, and the {\it Peiffer identity}, $\sft(h_1)\acton h_2= h_1h_2h_1^{-1}$, holds for all $g\in\sfG$ and $h,h_1,h_2\in\sfH$. Furthermore, a {\it differential crossed module} is the 1-jet of a Lie crossed module (see Section \ref{sec:GroupoidsLinfty}), and is given by a pair of Lie algebras $(\frg,\frh)$ with $\frg:=\sfLie(\sfG)$ and $\frh:=\sfLie(\sfH)$ with $\sft_*:\frh\to\frg$ such that $\sft_*(V\acton_* U)=[V,\sft_*(U)]$ and  $\sft_*(U_1)\acton_* U_2= [U_1,U_2]$ for all $V\in\frg$ and $U,U_1,U_2\in\frh$ where $\sft_*$ and $\acton_*$ are the linearisations of $\sft$ and $\acton$, respectively.\footnote{Differential crossed modules and 2-term $L_\infty$-algebras $(\sfL,\mu_i)$ with $\sfL=\sfL_{-1}\oplus\sfL_0$ and $\mu_3=0$ are actually the same thing. Indeed, given such an $L_\infty$-algebra, the corresponding differential crossed module is $\frg:=\sfL_0$ and $\frh:=\sfL_{-1}$, $\sft_*:=\mu_1$, $V\acton U:=\mu_2(U,V)$, $[U_1,U_2]:=\mu_2(\mu_1(U_1),U_2)$, and $[V_1,V_2]:=\mu_2(V_1,V_2)$ for $U,U_1,U_2\in\frh$ and  $V,V_1,V_2\in\frg$. The antisymmetry and the Jacobi identities for the Lie brackets $[-,-]$ as well as the equivariance condition $\sft_*(V\acton_* U)=[V,\sft_*(U)]$ follow from the higher Jacobi identities for $\mu_1$ and $\mu_2$, and the Peiffer condition $\sft_*(U_1)\acton_* U_2= [U_1,U_2]$ is evidently satisfied. Obviously, the converse is also true, i.e.~we can use the same identifications to construct a 2-term $L_\infty$-algebra $(\sfL,\mu_i)$ with $\sfL=\sfL_{-1}\oplus\sfL_0$ and $\mu_3=0$ from a differential crossed module, and the graded antisymmetry as well as the higher Jacobi identities for $\mu_1$ and $\mu_2$ follow from the Jacobi identities for the Lie brackets together with the equivariance and Peiffer conditions. }

A Deligne cocycle in the crossed module language is then given by
\begin{equation}
\{g_{ab},h_{abc},\lambda_{ab}, A_a,B_b\}
\end{equation}
with $g_{ab}:U_a\cap U_b\to\sfG$, $h_{abc}:U_a\cap U_b\cap U_c\to\sfH$, $\lambda_{ab}\in\Omega^1(U_a\cap U_b,\frh)$, $A_a\in\Omega^1(U_a,\frg)$, and $B_a\in\Omega^2(U_a,\frh)$ subject to the cocycle conditions
\begin{equation}
\begin{aligned}
 \sft(h_{abc})g_{ab}g_{bc}&=g_{ac}~,\\
 h_{acd}h_{abc}&=h_{abd}(g_{ab}\acton h_{bcd})~,\\
  \lambda_{ac}&=\lambda_{bc}+g_{bc}^{-1}\acton\lambda_{ab}-g_{ac}^{-1}\acton(h_{abc}\nabla_ah_{abc}^{-1})~,\\
 A_b&= g^{-1}_{ab} A_a g_{ab}+g^{-1}_{ab} \dd g_{ab}-\sft_*(\lambda_{ab})~,\\
 B_b&= g^{-1}_{ab}\acton B_a -\nabla_b \lambda_{ab}-\tfrac12[\lambda_{ab},\lambda_{ab}]~
 \end{aligned}
\end{equation}
on appropriate non-empty overlaps and $\nabla_a:=\dd+A_a\acton_*$. Furthermore, two such cocycles $\{g_{ab},h_{abc},\lambda_{ab}, A_a,B_b\}$ and $\{\tilde g_{ab},\tilde h_{abc},\tilde \lambda_{ab}, \tilde A_a,\tilde B_b\}$ whenever there is a coboundary transformation, mediated by
\begin{equation}
\{g_{a},h_{ab},\lambda_{a}\}
\end{equation}
with $g_{a}:U_a\to\sfG$, $h_{ab}:U_a\cap U_b\to\sfH$, and $\lambda_{a}\in\Omega^1(U_a,\frh)$, and explicitly given by
\begin{equation}
\begin{aligned}
\sft(h_{ab})g_{ab}g_b&=g_a \tilde g_{ab}\\
h_{ac}h_{abc}&=(g_a\acton \tilde h_{abc})h_{ab}(g_{ab}\acton h_{bc})~,\\
  \lambda_{a}&=\tilde\lambda_{ab}+\lambda_{b}+g_{b}^{-1}\acton\lambda_{ab}-g_{a}^{-1}\acton(h_{ab}\nabla_ah_{ab}^{-1})~,\\
\tilde{A}_a&=g_a^{-1} A_a g_a+g_a^{-1} \dd g_a-\sft_*(\lambda_a)~,\\
\tilde{B}_a&=g_a^{-1}\acton B_a -\tilde{\nabla}_a\Lambda_a-\tfrac12[\Lambda_a, \Lambda_a]~.\\
\end{aligned}
\end{equation}
It is rather straighforward to see that with the help of these coboundary transformations, we can always set $h_{aaa}=\unit_\sfH$, which, in turn, yields $g_{aa}=\unit_\sfG$ and $h_{aab}=h_{abb}=\unit_\sfH$. Residual coboundary transformations are then those with $h_{aa}=\unit_\sfH$.

\section{Homotopy Maurer--Cartan theory}

\subsection{Homotopy Maurer--Cartan equation and action}\label{sec:HMCEA}

Let $(\sfL,\mu_i)$ be an $L_\infty$-algebra. An element $a\in\sfL_1$ is called a {\it gauge potential}. We define its {\it curvature} $f\in\sfL_2$ by 
\begin{equation}\label{eq:curvature}
f:=\sum_{i\geq 1}\frac{1}{i!}\mu_i(a,\ldots,a)~.
\end{equation}
As is easily seen, due to the higher Jacobi identities~\eqref{eq:HJI}, the curvature satisfies the {\it Bianchi identity}
\begin{equation}\label{eq:Bianchi}
 \sum_{i\geq0}\frac{1}{i!}\mu_{i+1}(a,\ldots,a,f)= 0~.
\end{equation}

Furthermore, {\it gauge transformations} are mediated by elements $c_0\in\sfL_0$ and are given by
\begin{equation}
 \delta_{c_0}a:=\sum_{i\geq0} \frac{1}{i!}\mu_{i+1}(a,\ldots,a, c_0)~.
\end{equation}
Consequently,
\begin{equation}
\delta_{c_0}f=\sum_{i\geq 0}\frac{1}{i!}\mu_{i+2}(a,\ldots,a, f,c_0)~.
\end{equation}
Again using the higher Jacobi identities~\eqref{eq:HJI}, one can check that 
\begin{subequations}
\begin{equation}
  [\delta_{ c_0},\delta_{ c'_0}]a=\delta_{c''_0}a+\sum_{i\geq0}\frac{1}{i!}\mu_{i+3}(a,\ldots,a, f,c_0, c'_0)
\end{equation}
with
\begin{equation}
   c''_0:=\sum_{i\geq0}\frac{1}{i!}\mu_{i+2}(a,\ldots,a, c_0, c'_0)~.
\end{equation}
\end{subequations}
Hence, if 
\begin{equation}
 f=0
\end{equation}
gauge transformations do close.\footnote{Note that for $1$-term $L_\infty$-algebras they always close since there are no $\mu_i$ with $i\geq3$.} This equation is called the {\it homotopy Maurer--Cartan equation}, and solutions $a\in\sfL_1$ satisfying this equation are known as {\it Maurer--Cartan elements}.

The gauge parameters $c_0\in\sfL_0$ enjoy, in general, a gauge freedom mediated by next-to-lowest gauge parameters $c_{-1}\in\sfL_{-1}$. Likewise, the next-to-lowest gauge parameters $c_{-1}\in\sfL_{-1}$ enjoy, in general, a gauge freedom mediated by next-to-next-to-lowest gauge parameters $c_{-2}\in\sfL_{-2}$, and so on. These are the {\it higher gauge transformations} which are given by
\begin{equation}
\delta_{c_{-k-1}}  c_{-k}:=\sum_{i\geq0} \frac{1}{i!}\mu_{i+1}(a,\ldots,a, c_{-k-1})~,
\end{equation}
with $ c_{-k}\in\sfL_{-k}$. As one may check, if $f=0$, also the higher gauge transformations close.

Provided $(\sfL,\mu_i,\langle-,-\rangle)$ is a cyclic $L_\infty$-algebra with an inner product $\langle-,-\rangle$ of degree~$-3$, the homotopy Maurer--Cartan equation is variational. Indeed, $f=0$ follows from varying the gauge invariant action functional
\begin{equation}\label{eq:MCA}
 S_{\rm MC}:=\sum_{i\geq1} \frac{1}{(i+1)!}\langle a,\mu_i(a,\ldots,a)\rangle~.
\end{equation}

\subsection{$L_\infty$-morphisms revisited}

Let us now consider how Maurer--Cartan elements behave under  $L_\infty$-morphisms. To this end, let $(\sfL,\mu_i)$ and  $(\sfL',\mu'_i)$ be two $L_\infty$-algebras related by an $L_\infty$-morphism~\eqref{eq:LMor}. Under such a morphism, the gauge potential transforms according to
\begin{equation}\label{eq:LMorA}
 \sfL_1\ni a\mapsto a':=\sum_{i\geq 1} \frac{1}{i!}\phi_i(a,\dots,a)\in\sfL'_1~.
\end{equation}
Correspondingly, the curvatures~\eqref{eq:curvature} are related as
\begin{equation}
  \sfL_2\ni f\mapsto f'=\sum_{i\geq 0}\frac{1}{i!}\phi_{i+1}(a,\ldots,a,f)\in\sfL'_2~.
\end{equation}
Consequently, Maurer--Cartan elements are mapped to Maurer--Cartan elements under $L_\infty$-morphisms. 

In addition, a gauge transformation $a\mapsto a+\delta_{ c_0}a$ with gauge parameter $c_0$ of a Maurer--Cartan element $a\in\sfL_1$ is transformed under an $L_\infty$-morphism to  $a'\mapsto a'+\delta_{ c'_0}a'$ with $a'$ given by~\eqref{eq:LMorA} and 
\begin{equation}
 \sfL_0\ni c_0\mapsto c_0':=\sum_{i\geq 0} \frac{1}{i!}\phi_{i+1}(a,\dots,a, c_0)\in\sfL'_0~.
\end{equation}
Hence, gauge equivalence classes of Maurer--Cartan elements are mapped to gauge equivalence classes of Maurer--Cartan elements. 

The above can be extended so that for an $L_\infty$-quasi-isomorphism between two $L_\infty$-algebras  $(\sfL,\mu_i)$ and  $(\sfL',\mu'_i)$, the moduli space of Maurer--Cartan elements for $(\sfL,\mu_i)$  (i.e.~the space of solutions to the Maurer--Cartan equation modulo gauge transformations) is isomorphic to the moduli space of Maurer--Cartan elements for $(\sfL',\mu'_i)$.

\subsection{Higher Chern--Simons theory}

Recall from Section \ref{sec:Linfty} the tensor product $L_\infty$-algebra $\Omega^\bullet(X,\sfL)$ between the de Rham complex $(\Omega^\bullet(X),\dd,\int)$ on a $d$-dimensional compact oriented manifold $X$ without boundary for $d\geq3$ with a finite-dimensional cyclic $L_\infty$-algebra $(\sfL,\mu_i,\langle-,-\rangle)$ with $\sfL=\bigoplus_{k=-d+3}^0\sfL_k$. 

When $d=3$, $\sfL=\sfL_0$ and a gauge potential $a\in\Omega^\bullet_1(X,\sfL)$ is given by $A\in\Omega^1(X,\sfL_0)$. Correspondingly, the curvature $f\in\Omega^\bullet_2(X,\sfL)$ is given as $F:=\dd A+\tfrac12[A,A]\in\Omega^2(X,\sfL_0)$. Likewise, a gauge parameter $c_0\in\Omega^\bullet_0(X,\sfL)$ is given by an element $c\in\scrC^\infty(X,\sfL_0)$ and, consequently, gauge transformations $\delta_{c_0}a$ and $\delta_{c_0}f$ read as $\delta_c A=\dd c+[A,c]$ and $\delta_c F=-[c,F]$, respectively. The Maurer--Cartan action~\eqref{eq:MCA} then becomes
\begin{equation}
 S_{\rm MC}=\int_X\Big\{\tfrac12\langle A,\dd A\rangle+\tfrac{1}{3!}\langle A,[A,A]\rangle\Big\}\,,
\end{equation}
that is, we obtain {\it ordinary Chern--Simons theory}.

When $d=4$, we have $\sfL=\sfL_{-1}\oplus\sfL_0$ and a gauge potential $a\in\Omega^\bullet_1(X,\sfL)$ is given by $A+B\in\Omega^1(X,\sfL_0)\oplus\Omega^2(X,\sfL_{-1})$ and the curvature $f\in\Omega^\bullet_2(X,\sfL)$ is given by an element $F+H\in \Omega^2(X,\sfL_0)\oplus\Omega^3(X,\sfL_{-1})$ with
\begin{equation}\label{eq:CS4D}
\begin{aligned}
F&:=\dd A+\tfrac12\mu_2(A,A)+\mu_1(B)~,\\
H&:=\dd B+\mu_2(A,B)-\tfrac{1}{3!}\mu_3(A,A,A)~.
\end{aligned}
\end{equation}
Furthermore, a gauge parameter $c_0\in\Omega^\bullet_0(X,\sfL)$ is given by an element $c+\lambda\in\scrC^\infty(X,\sfL_0)\oplus\Omega^1(X,\sfL_{-1})$ and so, gauge transformations $\delta_{c_0}a$ and $\delta_{c_0}f$ read as
\begin{equation}
\begin{aligned}
 \delta_{c,\lambda}A&=\mbox{d}c+\mu_2(A,c)+\mu_1(\lambda)~,\\
 \delta_{c,\lambda}B&=-\mu_2(c,B)+\mbox{d}\lambda+\mu_2(A,\lambda)+\tfrac12\mu_3(c,A,A)~,\\
 \delta_{c,\lambda}F&=-\mu_2(c,F)~,\\
 \delta_{c,\lambda}H&=-\mu_2(c,H)+\mu_2(F,\lambda)-\mu_3(F,A,c)~.
\end{aligned}
\end{equation}
Finally, the Maurer--Cartan action~\eqref{eq:MCA} reads in this case as
\begin{equation}
\begin{aligned}
S_{\rm MC}&=\int_X\Big\{\langle B,\mbox{d} A+\tfrac12\mu_2(A,A)+\tfrac12\mu_1(B)\rangle\,+\\
&\kern1.5cm+\tfrac{1}{4!}\langle\mu_3(A,A,A), A\rangle\Big\}~.
\end{aligned}
\end{equation}
This is an instance of {\it higher Chern--Simons theory}. It is clear how this generalises to any dimension $d>4$. Note that this can also be generalised to Calabi--Yau manifolds to define {\it higher holomorphic Chern--Simons theory}~\cite{Saemann:2017vuy}: instead of using the de Rham complex one works with the Dolbeault complex and to define an action one uses the holomorphic measure. We shall come back to this in Section~\ref{sec:ambitwistorYM}. 

\subsection{Batalin--Vilkovisky formalism}

Let us now discuss the {\it Batalin--Vilkovisky formalism}~\cite{Batalin:1977pb,Batalin:1981jr,Batalin:1984jr,Batalin:1984ss,Batalin:1985qj} adapted to the context of $L_\infty$-algebras and the Maurer--Cartan action~\eqref{eq:MCA}.

To this end, let $(\sfL,\mu_i,\langle-,-\rangle)$ be an  $L_\infty$-algebra. In Section \ref{sec:Linfty}, we introduced the coordinate functions $\xi^\alpha$ with $|\xi^\alpha|\in\IZ$ on $\sfL[1]$ and the basis vectors $\tau_\alpha$  with $|\tau^\alpha|=-|\xi_\alpha|+1$ on $\sfL$. It is  convenient to define the contracted coordinate functions $\xi:=\xi^\alpha\tau_\alpha$ with total degree $|\xi|=1$. Effectively, we are considering $\sfL':=\scrC^\infty(\sfL[1])\otimes\sfL$, and this $\IZ$-graded vector space can be given an $L_\infty$-structure by 
\begin{equation}
\begin{aligned}
 \mu'_1(\zeta \otimes \ell) &:= (-1)^{|\zeta|}\zeta\otimes \mu_1(\ell)~,\\
  \mu'_i(\zeta_1 \otimes \ell_1,\ldots, \zeta_i \otimes \ell_i) &:= (-1)^{i\sum_{j=1}^i|\zeta_i|+\sum_{j=2}^{i} |\zeta_j|\sum_{k=1}^{j-1} |\ell_k|}\,\times\\
  &\kern1cm\times(\zeta_1\cdots \zeta_i)\otimes\mu_i(\ell_1,\ldots, \ell_i)~,
\end{aligned}
\end{equation}
and we shall refer to $|\zeta|\in\IZ$ as the {\it ghost degree}~\cite{Jurco:2018sby}. In this formulation, the action of the homological vector field~\eqref{eq:QVectorGen} is simply
\begin{equation}
Q\xi=-\sum_{i\geq1}\frac{1}{i!}\mu'_i(\xi,\ldots,\xi)~.
\end{equation}
Then, 
\begin{equation}
Q^2\xi=-\sum_{i\geq0,\,j\geq1}\frac{(-1)^i}{i!j!}\mu_{i+1}(\mu'_j(\xi,\ldots,\xi),\xi,\ldots,\xi)\ =\ 0
\end{equation}
by virtue of the Bianchi identity~\eqref{eq:Bianchi}. If, in addition, $\langle-,-\rangle$ is an inner product on $\sfL$ of degree $k$, then we can make $\sfL'$ cyclic by setting
\begin{equation}\label{eq:BFIP}
 \langle\zeta_1\otimes \ell_1,\zeta_2\otimes \ell_2\rangle':= 
(-1)^{k(|\zeta_1|+|\zeta_2|)+|\ell_1| |\zeta_2|}(\zeta_1\zeta_2)\langle\ell_1,\ell_2\rangle~.
\end{equation}

To BRST quantise the Maurer--Cartan action~\eqref{eq:MCA}, it is evident that we need to introduce ghosts due to the gauge invariance of the action. Moreover, due to the higher gauge redundancy, we also need to introduce higher ghosts i.e.~ghosts-for-ghost, ghosts-for-ghost-for-ghosts, etc. In particular, we need

\begin{center}
\begin{tabular}{|l|c|c|c|c|c|c|}
\hline
 & $a$ & $c_0$ & $c_{-1}$ & $\cdots$ & $c_{-k}$ & $\cdots$ \\ \hline
$L_\infty$-degree & $1$ & $0$ & $-1$ & $\dots$ & $-k$& $\cdots$ \\
ghost degree & 0 & 1 & 2 & $\cdots$ & $k+1$ & $\cdots$ \\
field type & b & f & b & $\cdots$ & f/b & $\cdots$ \\
\hline
\end{tabular}
\end{center}

\noindent
where `f' stands for {\it fermion} and `b' for {\it boson}, respectively. Thus, the {\it BRST field space} is 
\begin{equation}
\frF_{\rm BRST}:=\sfL_{\rm red}[1]\ewith \sfL_{\rm red}:=\bigoplus_{k\leq1}\sfL_k~.
\end{equation}
Inspired by our above discussion, we set
\begin{equation}
\sfL'_{\rm red}:=\scrC^\infty(\sfL_{\rm red}[1])\otimes\sfL_{\rm red}
\end{equation}
and use 
\begin{equation}
\sfa_{\rm red}:=a+\sum_{k\geq0}c_{-k}
\end{equation}
so that
\begin{equation}
Q_{\rm BRST}\sfa_{\rm red}=-\sum_{i\geq1}\frac{1}{i!}\mu'_i(\sfa_{\rm red},\ldots,\sfa_{\rm red})=-\sff_{\rm red}~.
\end{equation}
As we essentially truncated an $L_\infty$-algebra, $\sfL'_{\rm red}$ is, in general, {\it not} an $L_\infty$-algebra and thus, we do not expect to get $Q_{\rm BRST}^2=0$. Indeed, it is a straightforward but lengthy exercise to show that generically we have
\begin{equation}
Q_{\rm BRST}^2=0~~\mbox{mod}~~f=0~,
\end{equation}
where $f$ is the curvature of $a$. This is due to the fact that gauge transformations close generically only on-shell, see Section \ref{sec:HMCEA} which is known as {\it open symmetries} in the physics literature.

It is now obvious as how to cure this problem. We simply consider the whole of $\sfL$ thus effectively doubling the field content. Hence, in addition to the above fields, we also have

\begin{center}
\begin{tabular}{|l|c|c|c|c|c|c|c|}
\hline
 & $\cdots$ & $c^+_{-k}$ & $\cdots$ & $c^+_{-1}$ & $c_0^+$ &  $a^+$  \\ \hline
$L_\infty$-degree & $\cdots$ & $3+k$ & $\cdots$ & $4$ & $3$ & $2$  \\
ghost degree & $\cdots$ & $-k-2$ & $\cdots$ & $-3$ & $-2$ & $-1$ \\
field type & $\cdots$ & f/b & $\cdots$ & f & b & f  \\
\hline
\end{tabular}
\end{center}

\noindent
and which are known as {\it anti-fields}. This is known as the {\it Batalin--Vilkovisky formalism}. In particular, the {\it Batalin--Vilkovisky field space} is
\begin{equation}
\frF_{\rm BV}:=\sfL[1]\cong T^*[-1]\frF_{\rm BRST}~.
\end{equation}
Therefore,
\begin{equation}\label{eq:BVL}
\sfL':=\scrC^\infty(\sfL[1])\otimes\sfL
\end{equation}
so that
\begin{equation}
\sfa:=a+a^++\sum_{k\geq0}(c_{-k}+c_k^+)
\end{equation}
and
\begin{equation}
Q_{\rm BV}\sfa=-\sum_{i\geq1}\frac{1}{i!}\mu'_i(\sfa,\ldots,\sfa)=-\sff\quad\Longrightarrow\quad Q^2_{\rm BV}=0~.
\end{equation}
Furthermore, $\frF_{\rm BV}$ comes with a natural symplectic structure of degree $-1$ given by
\begin{equation}
 \omega_{\rm BV}:=-\tfrac12\langle\dd\sfa,\dd\sfa\rangle'
\end{equation}
with $\langle-,-\rangle'$ given in~\eqref{eq:BFIP}. In addition, letting $\{-,-\}_{\rm BV}$ be the Poisson bracket induced by $\omega_{\rm BV}$ and defining the {\it Maurer--Cartan--Batalin--Vilkovisky action}~\cite{Jurco:2018sby}
\begin{equation}\label{eq:BVA}
S_{\rm BV}:=\sum_{i\geq 1}\frac{1}{(i+1)!}\langle\sfa,\mu'_i(\sfa,\ldots,\sfa)\rangle'
\end{equation}
then
\begin{equation}\label{eq:QSBV}
 Q_{\rm BV}=\{S_{\rm BV},-\}_{\rm BV}
 \end{equation}
 with the nil-potency $Q^2_{\rm BV}=0$ being equivalent to the {\it classical master equation}
 \begin{equation}
\{S_{\rm BV},S_{\rm BV}\}_{\rm BV}=0~.
 \end{equation}
Notice that $\{S_{\rm BV},S_{\rm BV}\}_{\rm BV}=-\langle\sff,\sff\rangle'$ with the right-hand-side being identically zero for any $L_\infty$-algebra~\cite{Jurco:2018sby}.

\subsection{Yang--Mills theory}

It is evident from our above considerations that any variational theory comes with an underlying $L_\infty$-structure~\cite{Jurco:2018sby} that is encoded in the homological vector field $Q_{\rm BV}$. Furthermore, the action for such theory can be recast as a Maurer--Cartan--Batalin--Vilkovisky action~\eqref{eq:BVA}.

As a concrete example, let us consider Yang--Mills theory on a $4$-dimensional compact Riemannian manifold $X$ without boundary and with gauge Lie algebra $\frg$. We introduce the {\it second-order Yang--Mills complex} by setting~\cite{Movshev:2003ib,Movshev:2004aw,Zeitlin:2007vv,Zeitlin:2007yf}
\begin{subequations}
\begin{equation}\label{eq:YM2C}
\begin{aligned}
 &\underbrace{\Omega^0(X,\frg)}_{=:\,\sfL_0}\ \xrightarrow{~\mu_1\,:=\, \dd~}\  \underbrace{\Omega^1(X,\frg)}_{=:\,\sfL_1}\\
  &\kern1cm\xrightarrow{~\mu_1\,:=\, \dd\star\dd~}\  \underbrace{\Omega^{3}(X,\frg)}_{=:\,\sfL_2}\ \xrightarrow{~\mu_1\,:=\, \dd~}\  \underbrace{\Omega^4(X,\frg)}_{=:\,\sfL_3}~,
 \end{aligned}
\end{equation}
where $\star$ is the Hodge operator on $X$. This complex can be given an $L_\infty$-structure by defining the non-vanishing products by~\cite{Movshev:2003ib,Movshev:2004aw,Zeitlin:2007vv,Zeitlin:2007yf}
\begin{equation}
\begin{aligned}
\mu_1(c_1)&:=\dd c_1~,\\
\mu_1(A_1)&:=\dd{\star\dd A_1}~,\\
\mu_1(A^+_1)&:=\dd A^+_1~,\\
\mu_2(c_1,c_2)&:=[c_1,c_2]~,\\
\mu_2(c_1,A_1)&:=[c_1,A_1]~,\\ 
\mu_2(c_1,A^+_2)&:=[c_1,A^+_2]~,\\
\mu_2(c_1,c^+_2)&:=[c_1,c^+_2]~,\\ 
\mu_2(A_1,A^+_2)&:=[A_1,A^+_2]~,\\ 
\mu_2(A_1,A_2)&:=\dd{\star[A_1,A_2]}+[A_1,{\star\dd A_2}]+[A_2,{\star\dd A_1}]~,\\
\mu_3(A_1,A_2,A_3)&:=[A_1,\star[A_2,A_3]]\,+\\
 &\kern1cm+[A_2,\star[A_3,A_1]]+[A_3,\star[A_1,A_2]]
\end{aligned}
\end{equation}
\end{subequations}
for $c_{1,2}\in\sfL_0$, $A_{1,2,3}\in\sfL_1$, $A_2^+\in\sfL_2$, and $c_2^+\in\sfL_3$, respectively. This $L_\infty$-algebra can be made cyclic by
\begin{equation}\label{eq:YMIP}
 \langle \alpha_1\otimes t_1,\alpha_2\otimes t_2\rangle:=\int_X\alpha_1\wedge{\alpha_2}~\langle t_1,t_2\rangle~,
\end{equation}
where $\langle -,-\rangle$ on the right-hand-side is a metric on $\frg$. 

With these ingredients, it is a straightforward exercise to verify that the Maurer--Cartan--Batalin--Vilkovisky action~\eqref{eq:BVA} for the $L_\infty$-algebra~\eqref{eq:BVL} with $\sfL$ as given above yields
\begin{equation}\label{eq:YM2A}
 S_{\rm BV}=\int_X \Big\{\tfrac12 \langle F,{\star F}\rangle-\langle A^+,\nabla c\rangle+\tfrac12\langle c^+,[c,c]\rangle\Big\}
\end{equation}
with $F:=\dd A+\tfrac12[A,A]$ and $\nabla:=\dd +[A,-]$. This is simply the Batalin--Vilkovisky action for Yang--Mills theory~\cite{Batalin:1981jr}. The action of $Q_{\rm BV}$ is then given by
\begin{equation}
\begin{aligned}
 Q_{\rm BV}c &=-\tfrac12[c,c]~,\\
 Q_{\rm BV} A &= \nabla c~,\\
 Q_{\rm BV}A^+  &= -{\nabla{\star F}}-[c,A^+]~,\\
 Q_{\rm BV}c^+  &= {\nabla}{A^+}-[c,c^+]
 \end{aligned}
\end{equation}
as one can check using~\eqref{eq:QSBV}.

Yang--Mills theory in four dimensions admits an alternative formulation that only makes use of first-order and has only cubic interactions~\cite{Okubo:1979gt} which again can be formulated in $L_\infty$-language~\cite{Rocek:2017xsj,Jurco:2018sby}. In particular, consider the decomposition of differential $2$-forms 
\begin{equation}
\Omega^2(X,\frg)\cong\Omega^2_+(X,\frg)\oplus\Omega^2_-(X,\frg)
\end{equation}
into self-dual and anti-self-dual parts, respectively. We define the {\it first-order Yang--Mills complex}~\cite{Costello:2007ei}
\begin{subequations}
\begin{equation}
\begin{aligned}
 &\underbrace{\Omega^0(X,\frg)}_{=:\,\sfL_0}\ \xrightarrow{~\mu_1\,:=\,\dd~}\  \underbrace{\Omega^2_+(X,\frg)\oplus\Omega^1(X,\frg)}_{=:\,\sfL_1}\\
 &\kern1cm\xrightarrow{~\mu_1\,:=\,(\varepsilon+\dd)+P_+\dd~}\ \underbrace{\Omega^2_+(X,\frg)\oplus\Omega^3(X,\frg)}_{=:\,\sfL_2}\\
 &\kern2cm \xrightarrow{~\mu_1\,:=\,0+\dd~}\  \underbrace{\Omega^4(X,\frg)}_{=:\,\sfL_3}~,
  \end{aligned}
\end{equation}
where $P_+$ is the projector onto the self-dual $2$-forms and $\varepsilon\in\IR^+$. It can be augmented to a cyclic $L_\infty$-algebra by setting~\cite{Rocek:2017xsj,Jurco:2018sby}
\begin{equation}
\begin{aligned}
\mu_1(c_1)&:=\dd c_1~,\\
\mu_1(B_{+1}+A_1)&:=(\varepsilon B_{+1}+P_+\dd A_1)+\dd B_{+1}~,\\
\mu_1(A^+_1)&:=\dd A^+_1~,\\
\mu_2(c_1,c_2)&:=[c_1,c_2]~,\\
\mu_2(c_1,B_{+1}+A_1)&:=[c_1,B_{+1}]+[c,A_1]~,\\
\mu_2(c_1,B_{+1}^++A^+_1)&:=[c_1,B_{+1}^+]+[c,A^+_1]~,\\
\mu_2(c_1,c^+_2)&:=[c_1,c^+_2]~,\\ 
\mu_2(B_{+1}+A_1,B_{+2}+A_2)&:=P_+[A_1,A_2]\,+\\
&\kern1cm+[A_1,B_{+2}]+[A_2,B_{+1}]~,\\
\mu_2(B_{+1}+A_1,B_{+2}^++A^+_2)&:=[A_1,A_2^+]+[B_1,B_{+2}^+]~,
\end{aligned}
\end{equation}
\end{subequations}
where $c_i\in\sfL_0$, $(B_{+i}+A_i)\in\sfL_1$, $(B_{+i}^++A_i^+)\in\sfL_2$, and $c_i^+\in\sfL_3$ for $i=1,2$ together with the inner product~\eqref{eq:YMIP}.

The Maurer--Cartan--Batalin--Vilkovisky action~\eqref{eq:BVA} for the $L_\infty$-algebra~\eqref{eq:BVL} with this $\sfL$ then reads as
\begin{equation}\label{eq:YM1A}
\begin{aligned}
 S_{\rm BV}&=\int_X\Big\{\langle F, B_+\rangle+\tfrac\varepsilon2\langle B_+, B_+\rangle\,-\\
 &\kern1cm-\langle A^+,{\nabla c}\rangle-\langle B^+_+,[B_+,c]\rangle+\tfrac12\langle c^+,{[c,c]}\rangle\Big\}
 \end{aligned}
\end{equation}
and the action of $Q_{\rm BV}$ is given by
\begin{equation}
\begin{aligned}
 Q_{\rm BV}c&=-\tfrac12[c,c]~,\\
 Q_{\rm BV} (B_++A)&=-[c,B_+]+\nabla c~,\\
 Q_{\rm BV}(B^+_++A^+)&=-(F_++\varepsilon B_++[c,B^+_+])-(\nabla B_++[c,A^+])~,\\ 
 Q_{\rm BV}c^+&=\nabla A^++[B_+,B_+^+]-[c,c^+]~.
\end{aligned}
\end{equation}
Upon integrating out the fields $B_+$ and $B_+^+$, the first-order Yang--Mills action~\eqref{eq:YM1A} is the same as the second-order Yang--Mills action~\eqref{eq:YM2A} plus a topological term $\int_X\langle F,F\rangle$~\cite{Costello:2011aa,Jurco:2018sby}. Importantly, the $L_\infty$-algebras for the first-order and second-order formulations are, in fact, $L_\infty$-quasi-isomorphic~\cite{Rocek:2017xsj,Jurco:2018sby}.

\subsection{Interpretation of the Batalin--Vilkovisky $L_\infty$-algebra}

Upon inspecting the second-order Yang--Mills complex \eqref{eq:YM2C}, we realise that $\sfL_0$ encodes the gauge parameters, $\sfL_1$ the fundamental fields, and $\sfL_2$ the equations of motion. Moreover, the vector space $\sfL_3$ encodes all conserved currents (i.e. co-closed $1$-forms) as can be immediately seen by using the equivalent complex
\begin{equation}
\begin{aligned}
 &\underbrace{\Omega^0(X,\frg)}_{=:\,\tilde\sfL_0}\ \xrightarrow{~\mu_1\,:=\, \dd~}\  \underbrace{\Omega^1(X,\frg)}_{=:\,\tilde\sfL_1}\\
  &\kern1cm\xrightarrow{~\mu_1\,:=\, \dd^\dagger\dd~}\  \underbrace{\Omega^{1}(X,\frg)}_{=:\,\tilde\sfL_2}\ \xrightarrow{~\mu_1\,:=\, \dd^\dagger~}\  \underbrace{\Omega^0(X,\frg)}_{=:\,\tilde\sfL_3}~,
 \end{aligned}
\end{equation}
where $\dd^\dagger$ is the standard adjoint of $\dd$. 

In general, the $L_\infty$-algebra underlying a classical field theory has the following interpretation:
\begin{equation}
\begin{aligned}
 &\underbrace{\begin{array}{c}{\rm gauge}\\{\rm symmetries}\end{array}}_{\dots,~\sfL_{-1},~ \sfL_0}\ \longrightarrow \ \underbrace{\begin{array}{c}{\rm classical}\\{\rm fields}\end{array}}_{\sfL_1}\ \longrightarrow \\
  &\kern1cm\longrightarrow \ \underbrace{\begin{array}{c}{\rm equations}\\{\rm of~motion}\end{array}}_{\sfL_2}\ \longrightarrow \ \underbrace{\begin{array}{c}{\rm Noether}\\{\rm identities}\end{array}}_{\sfL_3,~\sfL_4,~\dots}
 \end{aligned}
\end{equation}

\section{Twistors and field theories}

Twistors~\cite{Penrose:1967wn} have been playing a fundamental role in the exploration of gauge and gravity theories as well as string theories. For instance, as an extension of encoding solutions to linear field equations in four dimensions in terms of cohomology groups on Penrose's twistor space by means of the {\it Penrose transform}~\cite{Penrose:1967wn,Penrose:1968me,Penrose:1969aa,Penrose:1972ia}, Ward~\cite{Ward:1977ta} (see also~\cite{Atiyah:1977pw}) proved that all solutions to the non-linear self-dual Yang--Mills equation on flat space-time have a natural interpretation in terms of holomorphic principal bundles over Penrose's twistor space. One often refers to this approach as the {\it Penrose--Ward transform}. This was generalised to the curved setting in~\cite{Atiyah:1978wi} (see also \cite{Penrose:1976js,Ward:1980am}). For detailed expositions on twistor theory and its applications see, for example, the text books~\cite{Manin:1988ds,Ward:1990vs,Mason:1991rf,Dunajski:2009aa} or the recent reviews~\cite{Wolf:2010av,Adamo:2011pv,Atiyah:2017erd,Adamo:2017qyl}. We shall now explain how the ideas twistor geometry can be combined with those of higher geometry to formulate higher gauge theories.

\subsection{A $6$-dimensional twistor space}

For the sake of concreteness, let us discuss the twistor space of \cite{0198535651,Saemann:2011nb,Mason:2011nw} that is associated with flat $6$-dimensional complexified space-time $M:=\IC^6$.\footnote{Reality conditions (to obtain e.g.~Minkowskian signature) can be imposed at any stage of the constructions. See~\cite{Saemann:2011nb} for details.}

In particular, the spin bundle on $M$ decomposes into the direct sum $S\oplus\tilde S$ of chiral and anti-chiral spinors leading to the identifications $TM\cong S\wedge S\cong\tilde S\wedge\tilde S$. We shall use $A,B,\ldots=1,\ldots,4$ to denote the chiral spinor indices, and because of these identifications, we may coordinatise $M$ by 
\begin{equation}
x^{AB}=-x^{BA}=\tfrac12\varepsilon^{ABCD}x_{CD}~,
\end{equation}
where $\varepsilon^{ABCD}$ is the Levi-Civita symbol in four dimensions. The next step is to consider the projectivisation $F:=\IP(S^*)\cong M\times\IP^3$, often called the {\it correspondence space}, which we equip with coordinates $(x^{AB},\lambda_A)$ with $\lambda_A$ being homogeneous coordinates on $\IP^3$. The correspondence space carries a natural rank-$3$ distribution, called the {\it twistor distribution}, generated by the vector fields $V^A=\lambda_B\partial^{AB}$ with $\partial^{AB}=\frac12\varepsilon^{ABCD}\partial_{CD}$ and $\partial_{AB}:=\frac{\partial}{\partial x^{AB}}$. Since the vector fields $V^A$ commute, the distribution they generate is integrable, and the corresponding $6$-dimensional leaf space is denoted by $P$ and called the {\it twistor space}. We thus have established the double fibration
\begin{equation}
\kern4pt \begin{picture}(50,40)
  \put(0.0,0.0){\makebox(0,0)[c]{$P$}}
  \put(64.0,0.0){\makebox(0,0)[c]{$M$}}
  \put(34.0,33.0){\makebox(0,0)[c]{$F$}}
  \put(7.0,18.0){\makebox(0,0)[c]{$\pi_1$}}
  \put(55.0,18.0){\makebox(0,0)[c]{$\pi_2$}}
  \put(25.0,25.0){\vector(-1,-1){18}}
  \put(37.0,25.0){\vector(1,-1){18}}
 \end{picture}
\end{equation}
Here, $\pi_2$ is the trivial projection. The projection $\pi_1$ is given by
\begin{equation}\label{eq:PIR}
\pi_1\,:\,(x^{AB},\lambda_A)\mapsto (x^{AB}\lambda_B,\lambda_A)
\end{equation}
and hence, the twistor space $P$ can be equipped with coordinates $(z^A,\lambda_A)$ subject to the constraint
\begin{equation}
z^A\lambda_A=0~.
\end{equation}
Because of this constraint, $P$ can be viewed as a quadric hypersurface in $\scrO_{\IP^3}(1)\otimes\IC^4\to\IP^3$.

The projection~\eqref{eq:PIR} is a $6$-dimensional generalisation of the {\it Penrose incidence relation}. By virtue of this relation, it is straightforward to realise that a point $x\in M$ in space-time corresponds to a submanifold $\pi_1(\pi_2^{-1}(x))\hookrightarrow P$ biholomorphic to $\IP^3$ in twistor space. Conversely, a point $(z,\lambda)\in P$ in twistor space corresponds to a submanifold $\pi_2(\pi_1^{-1}(z,\lambda))\hookrightarrow M$ in space-time given by
\begin{equation}
x^{AB}=x^{AB}_0+\varepsilon^{ABCD}\mu_C\lambda_D~.
\end{equation}
Here, $x^{AB}_0$ is a particular solution to $z^A=x^{AB}\lambda_B$ and $\varepsilon^{ABCD}\mu_C\lambda_D$ represents the homogeneous solution that is parametrised by three parameters $\mu_A$.\footnote{Note that $\mu_A$ cannot be proportional to $\lambda_A$.} Hence, the submanifold $\pi_2(\pi_1^{-1}(z,\lambda))\hookrightarrow M$ is a totally null 3-plane in $M$.

As explained in detail in~\cite{Saemann:2011nb}, the twistor space $P$ admits various dimensional reductions. In particular, upon reducing to four space-time dimensions, the twistor space $P$ can be reduced to the Penrose twistor space, the space of all totally null 2-planes in four dimensions, to the ambitwistor space, the space of all null rays in four dimensions, and the hyperplane twistor space, the space of all hyperplanes in four dimensions. As already mentioned, the Penrose twistor space plays a crucial role in the formulation  of chiral fields such as self-dual Yang--Mills fields \cite{Penrose:1976js,Ward:1977ta,Atiyah:1977pw,Atiyah:1978wi,Ward:1980am}. The ambitwistor space plays a key role in formulating full Yang--Mills theory~\cite{Witten:1978xx,Isenberg:1978kk,Isenberg:1978qd,Manin:1988ds,Buchdahl:1985aa,Pool:1981aa}, and, as shown in \cite{Saemann:2011nb,Saemann:2012uq}, the hyperplane twistor space is key to studying the self-dual string equation \cite{Howe:1997ue}.

\subsection{Zero-rest-mass fields}

As in four dimensions, also in six dimensions certain cohomology groups on twistor space encode the solutions to zero-rest-mast field equations. 

To define the notion of helicity in six dimensions, consider a null-vector $p$. The null-condition $p^2=0$ implies that $\det(p_{AB})=0=\det(p^{AB})$ so that
\begin{equation}
 p_{AB}\ =\  k_{Aa} k_{Bb}\varepsilon^{ab}\eand
 p^{AB}\ =\  \tilde{k}^{A\dot a} \tilde{k}^{B\dot b}\varepsilon_{\dot a\dot b}
\end{equation}
with $a,b,\ldots,\dot a,\dot b,\ldots=1,2$ and $\varepsilon^{ab}$ and $\varepsilon_{\dot a\dot b}$ being the $2$-dimensional Levi-Civita symbols. Evidently, the transformations $k_{Aa}\mapsto M\ {\!\!_a}^b  k_{Ab}$ and $\tilde{k}^{A\dot a}\mapsto \tilde{M}\ {\!\!^{\dot a}}_{\dot b} \tilde{k}^{A\dot b}$ with $\det M=1=\det \tilde{M}$ do not alter the momentum $p$ so that $a,\dot a,\ldots$ are, in fact, little group indices.  Consequently, the little group is $\sfSL(2,\IC)\times \widetilde{\sfSL(2,\IC)}$. It should be noted that $k_{Aa} \tilde{k}^{A\dot b}=0$ since $p_{AB}=\frac12\varepsilon_{ABCD}p^{CD}$. Chiral zero-rest-mass fields will transform trivially under $\widetilde{\sfSL(2,\IC)}$ and hence, they are characterised by the {\it helicity} $h\in\frac12\IN_0$. For instance, a $3$-form curvature $H=\dd B$ reads in spinor notation as $H=(H_{AB},H^{AB})=(\partial_{C(A}B_{B)}{}^C,\partial^{C(A}B_C{}^{B)})$ with $H_{AB}$ representing the self-dual part of $H$ and $H^{AB}$ the anti-self-dual part, respectively. Hence, imposing self-duality amounts to putting $H^{AB}=0$ and the three polarisation states of a helicity $1$ field $H_{AB}$ are then given as
\begin{equation}
 H_{AB\,ab}=k_{A(a}k_{Bb)}\de^{\di p\cdot x}~.
\end{equation}
See \cite{Cheung:2009dc,Saemann:2011nb,Mason:2011nw} for more details. 

Next, we define the sheaf $\scrZ_h$ of chiral rest-mass fields of helicity $h$ by
\begin{equation}
\begin{aligned}
\scrZ_{h=0}&:=\ker\Big\{\square:=\tfrac14 \partial^{AB}\partial_{AB}\,:\,\det(S^*) \to \otimes^2\det(S^*)\Big\}~,\\ 
 \scrZ_{h>0}&:=\ker\Big\{\partial^{AB}\,:\,(\odot^{2h}S^*)\otimes\det(S^*)\to\\
 &\kern2cm\to(\odot^{2h-1}S^*\otimes S)_0\otimes\otimes^2\det(S^*)\Big\}~. 
\end{aligned}
\end{equation}
The powers of the determinant of $S^*$ are included to render the zero-rest-mass field equations conformally invariant. As was proved in~\cite{0198535651,Saemann:2011nb,Mason:2011nw}, we have the identifications for any open convex subset $U\subseteq M$ 
\begin{equation}\label{eq:TISO}
 H^3(\hat U,\scrO_{\hat U}(-2h-4))\cong H^0(U,\scrZ_h)\cong H^2(\hat U,\scrO_{\hat U}(2h-2))~,
\end{equation}
where $\hat U:=\pi_1(\pi_2^{-1}(U))\subseteq P$. 

The first isomorphism is a direct generalisation of the Penrose transform, and it can be expressed in terms of contour integral formul{\ae} as
\begin{subequations}
\begin{equation}
 \phi_{A_1\cdots A_{2h}}(x)=\oint_\scrC \Omega^{(3,0)}~ \lambda_{A_1}\cdots\lambda_{A_{2h}} f_{-2h-4}(x\cdot\lambda,\lambda)
\end{equation}
for $ f_{-2h-4}$ a representative of $H^3(\hat U,\scrO_{\hat U}(-2h-4))$ and 
\begin{equation}\label{eq:Omega}
 \Omega^{(3,0)}:=\tfrac{1}{4!}\varepsilon^{ABCD}\lambda_A\dd\lambda_B\wedge\dd\lambda_C\wedge \dd \lambda_D~.
\end{equation}
\end{subequations}
It is easily checked that fields arising from such integral formul{\ae} satisfy the appropriate zero-rest-mass field equations. The second isomorphism in \eqref{eq:TISO} is a generalisation of the Penrose--Ward transform (in the Abelian setting). 

These two isomorphisms allow for a twistor space action for chiral zero-rest-mass fields~\cite{Saemann:2011nb,Mason:2011nw}. Indeed, the holomorphic measure on $P$ is a (6,0)-form of homogeneity $+6$ given by
\begin{equation}
 \Omega^{(6,0)}:=\oint_\scrC\frac{\Omega^{(4,0)}(z)\wedge \Omega^{(3,0)}(\lambda)}{z^A\lambda_A}~,
\end{equation}
where $\scrC$ is any contour encircling $P$ inside $\scrO_{\IP^3}(1)\otimes\IC^4\to\IP^3$, $\Omega^{(3,0)}(\lambda)$ given by~\eqref{eq:Omega}, and $\Omega^{(4,0)}(z)$ is 
$ \Omega^{(4,0)}(z):=\tfrac{1}{4!}\varepsilon_{ABCD}\dd z^A\wedge\dd z^B\wedge\dd z^C\wedge \dd z^D$. We then consider the twistor space action
\begin{equation}\label{eq:ATA}
 S:=\int_{\hat U}\Omega^{(6,0)}\wedge B^{(0,2)}\wedge \bar\partial C^{(0,3)}
\end{equation}
for the differential forms $C^{(0,3)}\in\Omega^{(0,3)}(\hat U,\scrO_{\hat U}(-2h-4))$ and $B^{(0,2)}\in\Omega^{(0,2)}(\hat U,\scrO_{\hat U}(2h-2))$. Hence, on-shell, we find $\bar\partial C^{(0,3)}=0=\bar\partial B^{(0,2)}$, and, consequently, by the {\v C}ech--Dolbeault correspondence, these differential forms correspond to representatives of the {\v C}ech cohomology groups $H^3(\hat U,\scrO_{\hat U}(-2h-4))$ and $H^2(\hat U,\scrO_{\hat U}(2h-2))$, respectively.

\subsection{Generalisations: non-Abelian fields}

Firstly, we would like to generalise the above to a non-Abelian setting. Helicity $h$ zero-rest-mass fields are described by $H^2(\hat U,\scrO_{\hat U}(2h-2))$, and for $h=1$, that is, a self-dual $3$-form curvature, we have  $H^2(\hat U,\scrO_{\hat U})$. To analyse this cohomology group, we consider the exponential sheaf sequence
\begin{equation}
0\to\IZ\to\scrO_{\hat U}\to \scrO^\times_{\hat U}\to0~.
\end{equation}
The corresponding induced long exact cohomology sequence then yields
\begin{equation}
\begin{aligned}
& H^1(\hat U,\scrO_{\hat U}^\times)\stackrel{c_1}{\longrightarrow} H^2(\hat U,\IZ)\longrightarrow H^2(\hat U,\scrO_{\hat U})\\
 &\kern1cm \longrightarrow H^2(\hat U,\scrO_{\hat U}^\times)\stackrel{\rm DD}{\longrightarrow} H^3(\hat U,\IZ)~,
 \end{aligned}
\end{equation}
where $c_1$ is the first Chern class and DD the Dixmier--Duady class. Here, $H^1(\hat U,\scrO_{\hat U}^\times)$ is the moduli space of holomorphic line bundles and $H^2(\hat U,\scrO_{\hat U}^\times)$ the moduli space of holomorphic gerbes over $\hat U$, respectively. Since $c_1$ is surjective and $H^3(\hat U,\IZ)=0$, we obtain the identification
\begin{equation}
H^2(\hat U,\scrO_{\hat U})\cong H^2(\hat U,\scrO_{\hat U}^\times)~.
\end{equation}
This means that a holomorphic gerbe becomes holomorphically trivial when restricted to $\pi_1(\pi_2^{-1}(x))\hookrightarrow P$ for all $x\in M$. In spirit of the $4$-dimensional case~\cite{Manin:1988ds}, we shall call this property {\it $M$-triviality}.

In~\cite{Saemann:2012uq,Saemann:2013pca,Jurco:2014mva,Jurco:2016qwv}, the cohomology group $H^2(\hat U,\scrO_{\hat U})$ and its identification with the moduli space of solutions to certain field equations was generalised  to the cohomology set of principal $\scrG$-bundles for $\scrG$ a Lie quasi -group. This, in turn, can be understood as a direct generalisation of the Penrose--Ward transform to higher principal bundles.\footnote{In fact, it has been generalised to Lie quasi-groupoids~\cite{Jurco:2016qwv}.} For concreteness, let $\scrG$ be a Lie $2$-quasi-group with the associated $L_\infty$-algebra $(\sfL,\mu_i)$. On $U\subseteq M$ we consider the equations
\begin{equation}
F=0\eand H=\star H
\end{equation}
with $H$ and $F$ given by~\eqref{eq:CS4D}. It was then shown in~\cite{Saemann:2012uq,Saemann:2013pca,Jurco:2014mva,Jurco:2016qwv} that the moduli space of solutions to these equations is equivalent to the moduli space of holomorphic principal $\scrG$-bundles over $\hat U\subseteq P$ which are $M$-trivial when restricted to $\pi_1(\pi_2^{-1}(x))\hookrightarrow P$ for all $x\in M$. 

The question as how to extend the twistor action~\eqref{eq:ATA} to this setting has remained open. Here, we would like to offer a solution. The {\v C}ech--Dolbeault correspondence extends to higher principal bundles~\cite{Saemann:2012uq,Saemann:2013pca,Jurco:2014mva,Jurco:2016qwv}. Consequently, a holomorphic principal $\scrG$-bundle for $\scrG$ a Lie $2$-quasi-group can be equivalently described by a complex principal $\scrG$-bundle equipped with a connective structure locally given by $A^{(0,1)}+B^{(0,2)}\in\Omega^{(0,1)}(\hat U,\sfL_0)\oplus\Omega^{(0,2)}(\hat U,\sfL_{-1})$ subject to the equations
\begin{subequations}\label{eq:HHCS6D}
\begin{equation}
F^{(0,2)}=0\eand H^{(0,3)}=0~,
\end{equation}
where
\begin{equation}
\begin{aligned}
F^{(0,2)}&:=\bar\partial A^{(0,1)}+\tfrac12\mu_2(A^{(0,1)},A^{(0,1)})+\mu_1(B^{(0,2)})~,\\
H^{(0,3)}&:=\bar\partial B^{(0,2)}+\mu_2(A^{(0,1)},B^{(0,2)})\,-\\
&\kern1.5cm-\tfrac{1}{3!}\mu_3(A^{(0,1)},A^{(0,1)},A^{(0,1)})~.
\end{aligned}
\end{equation}
\end{subequations}
The $M$-triviality is encoded in the assumptions of the existence of a gauge in which both $A^{(0,1)}$ and $B^{(0,2)}$ have no components along the submanifolds $\IP^3\hookrightarrow P$. To write down an action for these equations, let us also consider $C^{(0,3)}\in\Omega^{(0,3)}(\hat U,\scrO_{\hat U}(-6)\otimes\sfL_0)$ and $D^{(0,4)}\in\Omega^{(0,4)}(\hat U,\scrO_{\hat U}(-6)\otimes\sfL_{-1})$ and assume that $\sfL$ come equipped with a cyclic inner product $\langle-,-\rangle$. With these ingredients, the most general holomorphic higher Chern--Simons action we can write down is
\begin{equation}
\begin{aligned}
S&:=\int_{\hat U}\Omega^{(6,0)}\wedge\Big\{\langle B^{(0,2)},\bar\partial C^{(0,3)}\rangle+\langle D^{(0,4)},\bar\partial A^{(0,1)}\rangle\,+\\
&\kern1.5cm+\tfrac12\langle D^{(0,4)},\mu_2(A^{(0,1)},A^{(0,1)})\,+\\
&\kern1.5cm+\langle D^{(0,4)},\mu_1(B^{(0,2)})\rangle\,-\\
&\kern1.5cm-\langle\mu_2(A^{(0,1)},B^{(0,2)}),C^{(0,3)}\rangle\,+\\
&\kern1.5cm+\tfrac{1}{3!}\langle\mu_3(A^{(0,1)},A^{(0,1)},A^{(0,1)}), C^{(0,3)}\rangle\Big\}~.
\end{aligned}
\end{equation}
Evidently, this action reduces to~\eqref{eq:ATA} in the Abelian case.\footnote{Note that $H^1(\hat U,\scrO_{\hat U})=0$ and $H^4(\hat U,\scrO_{\hat U}(-6))=0$.} It also reproduces the equations~\eqref{eq:HHCS6D} plus some equations for $C^{(0,3)}$ and $D^{(0,4)}$ in the background of $A^{(0,1)}$ and $B^{(0,2)}$, respectively.

\subsection{Generalisations: supersymmetry}

As was shown in \cite{Chern:2009nt,Saemann:2012uq,Mason:2012va}, the twistor space $P$ admits an extension to accommodate $\calN=(n,0)$ supersymmetry. In particular, one replaces space-time by chiral superspace $M:=\IC^{6|8n}$ equipped with coordinates $(x^{AB},\eta^A_I)$ for $I,J,\ldots=1,\ldots,2n$. The supersymmetry algebra then is
\begin{subequations}
\begin{equation}
\{D^I_A,D^J_B\}=-4\Omega^{IJ}P_{AB}
\end{equation}
where $\Omega^{IJ}$ is an $\sfSp(n)$-invariant $(2n\times 2n)$-matrix and
\begin{equation}
D^I_A:=\der{\eta^A_I}-2\Omega^{IJ}\eta^B_J\der{x^{AB}}\eand P_{AB}:=\der{x^{AB}}~.
\end{equation}
\end{subequations}
The correspondence space then becomes $F:=\IC^{6|4n}\times\IP^3$ with coordinates  $(x^{AB},\eta^A_I,\lambda_A)$. The twistor distribution is now generated by the same bosonic vector fields $V^A:=\lambda_B P^{AB}$ together with the fermionic vector fields $V^{I\,AB}:=\frac12\varepsilon^{ABCD}\lambda_C D_D^I$, and it is of rank $3|6n$. The twistor space $P$ is again the leaf space obtained by quotienting $F$ by the twistor distribution and of dimension $6|2n$. It can be equipped with the coordinates $(z^A,\eta_I,\lambda_A)$ subject to quadric constraint
\begin{equation}
 z^A\lambda_A=\Omega^{IJ}\eta_I\eta_J~,
\end{equation}
and the Penrose incidence relations take the form
\begin{equation}
 z^A=(x^{AB}+\Omega^{IJ}\eta^A_I\eta^B_J)\lambda_B\eand \eta_I=\eta^A_I\lambda_A~.
\end{equation}

In~\cite{Saemann:2012uq,Saemann:2013pca,Jurco:2014mva,Jurco:2016qwv} it was proved, that the moduli space of $M$-trivial holomorphic principal $\scrG$-bundles, for $\scrG$ a Lie quasi-group, over this twistor space is naturally identified with the moduli space of solutions to the constraint system of supercurvatures containing the non-Abelian tensor multiplet. In fact, this identification is lifted to the level of an $L_\infty$-quasi-isomorphism.

\subsection{Yang--Mills theory}\label{sec:ambitwistorYM}

Finally, we would like to revisit $\calN=3$ supersymmetric Yang--Mills theory in four dimensions in the context of twistor theory.\footnote{See~\cite{Saemann:2012rr,Lechtenfeld:2012ye,Ivanova:2013vya} for a twistorial discussion of (maximally supersymmetric) Yang--Mills theory in six dimensions.}

It was shown in~\cite{Witten:1978xx,Isenberg:1978kk,Isenberg:1978qd}, that the moduli space of solutions to the constraint system of supercurvatures describing  $\calN=3$ supersymmetric Yang--Mills theory on $\calN=3$ superspace is naturally identified with the moduli space of $M$-trivial holomorphic principal $\sfG$-bundles, for $\sfG$ a Lie group, over ambitwistor space $L$. This constraint system is equivalent to the $\calN=3$ supersymmetric Yang--Mills equations on ordinary space-time \cite{Harnad:1984vk,Harnad:1985bc} which, in turn, are equivalent to the maximally supersymmetric Yang--Mills equations. The ambitwistor space in question is a supermanifold, and because of the peculiar choice of supersymmetry, a Calabi--Yau supermanifold~\cite{Witten:2003nn}. However, as the bosonic part of this ambitwistor space is $5$-dimensional, an action on ambitwistor space for $\calN=3$ supersymmetric Yang--Mills theory \`a la ordinary holomorphic Chern--Simons theory (via the {\v C}ech--Dolbeault correspondence) appears not possible. In~\cite{Saemann:2017vuy} a solution to this conundrum was proposed in terms of higher holomorphic Chern--Simons theory. It is important to note that the action proposed in~\cite{Saemann:2017vuy}  differs from an earlier proposal~\cite{Mason:2005kn} in that it makes solely use of the underlying complex geometry and works for any space-time signature.

In particular, the ambitwistor space $L$ is a $5|6$-dimen\-sional supermanifold and hence, a natural candidate to consider is higher holomorphic Chern--Simons theory for a Lie 3-quasi-group. Indeed, in this case the connective structure is given by a $(0,1)$-form $A^{(0,1)}$, a $(0,2)$-form $B^{(0,2)}$, and a $(0,3)$-form $C^{(0,3)}$. The $M$-triviality is encoded in the assumption of the existence of a gauge in which these differential forms have no components along certain submanifolds which in the case at hand is biholomorphic to $\IP^1\times\IP^1$. In addition to this, we shall work in a gauge~\cite{Witten:2003nn} in which these differential forms have only a holomorphic dependence on the fermionic coordinates and, in addition, have no anti-holomorphic fermionic directions. Under these assumptions, we may consider the action
\begin{equation}
\begin{aligned}
S&:=\int_L \Omega^{5|6,0}\wedge\Big\{\langle A^{0,1},\bar\partial C^{0,3}\rangle+\tfrac12\langle B^{0,2},\bar\partial B^{0,2}\rangle+\\
   &\kern1cm+\langle B^{0,2},\mu_1(C^{0,3})\rangle+\tfrac{1}{2} \langle A^{0,1},\mu_2(A^{0,1},C^{0,3})\rangle\,+\\
   &\kern1cm+\tfrac{1}{2}\langle A^{0,1},\mu_2(B^{0,2},B^{0,2})\rangle\,+\\
   &\kern1cm+\tfrac{1}{3!}\langle A^{0,1},\mu_3(A^{0,1},A^{0,1},B^{0,2})\rangle\,+\\
 &\kern1cm+\tfrac{1}{5!}\langle A^{0,1},\mu_4(A^{0,1},A^{0,1},A^{0,1},A^{0,1})\rangle\Big\}~,
\end{aligned}
\end{equation}
where $\Omega^{5|6,0}$ is the globally defined no-where vanishing holomorphic measure on ambitwistor space.  Here, the integration over the holomorphic fermionic directions has to be understood in the sense of Berezin. 

Upon varying this action, we find, for instance,
\begin{equation}
\bar\partial A^{(0,1)}+\tfrac12\mu_2(A^{(0,1)},A^{(0,1)})+\mu_1(B^{(0,2)})=0~.
\end{equation}
Thus, transitioning to the minimal model as discussed in Section~\ref{sec:Linfty}, we recover the equations which are equivalent to the constraint system of $\calN=3$  supersymmetric Yang--Mills theory~\cite{Witten:1978xx,Isenberg:1978kk,Isenberg:1978qd}.

\bibliographystyle{prop2015}
\bibliography{allbibtex}

\end{document}